\documentclass[a4paper,12pt]{article}

\usepackage[pages=all, color=black, position={current page.south}, placement=bottom, scale=1, opacity=1, vshift=5mm]{background}
\SetBgContents{
	\tt This work is shared under a \href{https://creativecommons.org/licenses/by-sa/4.0/}{CC BY-SA 4.0 license} unless otherwise noted
}      

\usepackage[margin=1in]{geometry} 

\usepackage{amsmath}
\usepackage{amsthm}
\usepackage{mathtools}
\usepackage{amssymb}
\usepackage{bbm}
\usepackage{rotating}
\usepackage{pdflscape}
\usepackage{array,longtable}
\usepackage{xltabular}
\usepackage{soul}
\usepackage{textcomp} 
\usepackage{mathcomp} 
\usepackage{makecell}
\usepackage{algpseudocode}
\usepackage{algorithm} 

\newenvironment{varalgorithm}[1]
  {\algorithm[H]\renewcommand{\thealgorithm}{#1}}
  {\endalgorithm}

\usepackage[utf8]{inputenc}
\usepackage{hyperref}
\hypersetup{
	unicode,
	pdfauthor={Author One, Author Two, Author Three},
	pdftitle={A simple article template},
	pdfsubject={A simple article template},
	pdfkeywords={article, template, simple},
	pdfproducer={LaTeX},
	pdfcreator={pdflatex}
}

\usepackage[sort&compress,numbers,square]{natbib}
\newcommand{\argmin}{\arg\!\min}

\theoremstyle{plain}
\newtheorem{theorem}{Theorem}
\newtheorem{corollary}[theorem]{Corollary}

\theoremstyle{definition}
\newtheorem{definition}[theorem]{Definition}

\usepackage{graphicx, color}
\graphicspath{{fig/}}

\newcommand{\Var}{\mathrm{Var}}

\usepackage{algorithm, algpseudocode} 
\usepackage{mathrsfs} 

\newcommand{\argmax}{\mathop{\mathrm{argmax}}}  

\newcolumntype{L}[1]{>{\raggedright\let\newline\\\arraybackslash\hspace{0pt}}m{#1}}
\newcolumntype{C}[1]{>{\centering\let\newline\\\arraybackslash\hspace{0pt}}m{#1}}
\newcolumntype{R}[1]{>{\raggedleft\let\newline\\\arraybackslash\hspace{0pt}}m{#1}}

\title{Assessing Inf\hspace{+0.05em}luential Observations in Pain Prediction using fMRI Data}
\author{Dongliang Zhang$^1$ \and Masoud Asgharian$^2$ \and Martin A.~Lindquist$^1$}

\date{
	$^1$Department of Biostatistics, Johns Hopkins University, Baltimore, Maryland, U.S.A. \\  
	$^2$Department of Mathematics and Statistics, McGill University, Montr\'{e}al, Qu\'{e}bec, Canada \\ 
}

\begin{document}

\maketitle
	
\begin{abstract}
Neuroimaging data allows researchers to model the relationship between multivariate patterns of brain activity and outcomes related to mental states and behaviors. However, the existence of outlying participants can potentially undermine the generalizability of these models and jeopardize the validity of downstream statistical analysis. To date, the ability to detect and account for participants unduly influencing various model selection approaches have been sorely lacking. Motivated by a task-based functional magnetic resonance imaging (fMRI) study of thermal pain, we propose and establish the asymptotic distribution for a diagnostic measure applicable to a number of dif{}ferent model selectors. A high-dimensional clustering procedure is further combined with this measure to detect multiple inf\hspace{+0.05em}luential observations. In a series of simulations, our proposed method demonstrates clear advantages over existing methods in terms of improved detection performance, leading to enhanced predictive and variable selection outcomes. Application of our method to data from the thermal pain study illustrates the influence of outlying participants, in particular with regards to differences in activation between low and intense pain conditions. This allows for the selection of an interpretable model with high prediction power after removal of the detected observations. Though inspired by the fMRI-based thermal pain study, our methods are broadly applicable to other high-dimensional data types. 

\end{abstract}

\newpage 

\tableofcontents

\newpage 
	
\section{Introduction}
\label{sec:introduction}

Neuroimaging is poised to take a substantial leap forward in uncovering the neurophysiological underpinnings of human behavior. This is primarily being driven by the improved analytic and computational capabilities provided by artificial intelligence, machine learning, and associated statistical techniques. This promises to allow for the creation of complex brain models, using multivariate patterns of brain activity as input, that will allow a better understanding of the functional representations underlying behavior, performance, clinical status, and prognosis \citep{orru2012using, haynes2015primer}. To reach their full potential, it is critical that these models are robust and generalizable across individuals, with strong performance on new, out-of-sample participants. However, the existence of participants with unusual traits or data could unduly influence the formation of the resulting brain models. Therefore, when developing such approaches, it is critical to identify when this occurs to ensure the generalizability of the proposed statistical models. 

During the course of a task-based functional magnetic resonance imaging (fMRI) study, participants receive one or more stimuli while their brain activation is measured at hundreds of time points. The measurement at each time point consists of the participant's blood oxygenation level dependent (BOLD) response at a large number of spatial locations (voxels), thus giving rise to multivariate time series data.  To perform a proper statistical analysis at the population level, it is necessary for voxels to lie in the same location across participants. The preprocessing of fMRI data therefore includes using nonlinear transformations to warp individual participants’ anatomical data to a common reference space (e.g., ``Montreal Neurologic Institute'' (MNI) space). The transformations are then applied to the functional data, allowing for the direct comparison of data across participants \citep{lindquist2008statistical, ombao2016handbook}. Next, for each participant, a voxel-wise general linear model (GLM) analysis is performed and the estimated regression coefficient corresponding to the task-specific regressor at each voxel is combined into a single subject-specific 3D activation map. This data is often used as input features in a predictive model. For models to be generalizable it is important that the data is both comparable across participants and measures equivalent psychological or physical phenomena. As outlined below, these assumptions may not always hold in practice.

As a motivating example, consider the prediction of physical pain using brain activation maps obtained from fMRI data \citep{lindquist2017}. Pain is associated with large social and economic costs. However, it is primarily assessed via self-report, making it hard to accurately determine. Recently, predictive models have been used to derive patterns of activity across brain regions that provide direct measures of pain intensity \citep{wager2013fmri, lindquist2017}. However, to be clinically useful these models must exhibit high sensitivity and specificity to pain outcomes. Inter-subject differences in both features and outcomes would be problematic and could arise for a number of reasons. First, there may exist individual differences in how participants process pain that influence how different brain regions activate in response to a painful stimulus. For example, not all participants rate a thermal stimulation at a certain temperature as equally painful, and a certain participants may not even find that particular stimulus painful at all.  Second, though data from all participants are transformed to a common reference space prior to analysis, there may remain inter-subject variability in functional topology that causes brain locations to be misaligned across participants \citep{wang2022bayesian, wang2023improved}. This could be problematic as brain activation from different locations are used as features in predictive models, and variation in functional location across participants will lead to feature misalignment when training models. Thus, participants whose functional topology differs substantially from the norm, could unduly influence model performance. Third, fMRI data is inherently noisy and prone to artifacts \citep{lindquist2008statistical, ombao2016handbook}. Together these three factors can lead to the presence of multiple outliers that negatively affect the formation, performance and generalizability of a brain model. Indeed, models for accessing physical pain could be more susceptible to these issue, as increased pain could potentially cause larger motion related artifacts. This can give rise to drastic changes in BOLD signal, leading to substantial variations in imaging intensity. Thus, the existence of motion outliers in fMRI-based pain studies further impedes the development of a universally valid predictive model, leading to reduced inferential accuracy and reliability.

Together these issues point to the importance of detecting participants whose data significantly impact the performance of the predictive model. In statistics, the detection of participants with unusual characteristics or outlying measurements is known as influential diagnosis. This refers to the detection of an isolated subset of the data exhibiting disproportionate influence on various aspects of an estimated model \citep{belsleybook}. This subset of data is commonly referred to as influential (or contaminated) observations or data points. 

To identify influential observations, it is imperative to first recognize that their detection depends heavily on the particular statistical model being used. In many modern applications, including the fMRI example discussed above, influential points may directly impact model specification as the precise mathematical form of the population is often unspecified beforehand. Indeed, when the number of predictors $p$ substantially surpasses the sample size $n$, high-dimensional variable selection is a fundamental instrument in exploratory data analysis, designated to achieve data reduction and attain an interpretable submodel. A non-exhaustive list of such procedures includes the LASSO \citep{lasso}, scaled LASSO \citep{sun2012}, elastic net \citep{zou2005}, SCAD \citep{fan2001} and MCP \citep{zhang2010}.

To further promote discussion of this topic and gain tangible insights into the pitfalls of ignoring observations exerting disproportionate influence on variable selection, 
a synthetic numerical example is given in Section 1 of the Supplementary Materials Part I (Zhang, Asgharian and Lindquist (2026)). This illustrates the non-negligible empirical probability of selecting an incorrect submodel based on the full, contaminated dataset. Moreover, this example also explicitly exhibits the adverse impacts of neglecting contaminated observations on the downstream analysis including prediction and model coef{}ficient estimation. Specifically, the mean squared error (MSE) for the predicted responses, the correlation between predicted and true responses, and the MSE for coef{}ficient estimates are all negatively influenced by the presence of influential points. Therefore, given the importance and ubiquity of stochastic variable selection, the possible existence of numerous outliers affecting variable selection especially in fMRI-based pain studies, and the complications caused by ignoring them, it is necessary to identify points influencing the data-dependent choice of a statistical submodel. 

Despite its importance, studies on influential diagnostics in model and variable selection is relatively scarce. In addition, to the best of our knowledge there are currently no methods of this type regularly used in neuroimaging. 
In the low-dimensional setting, \cite{altman1993} pioneered the study by introducing a variant of Cook's distance \citep{cook1977} incorporating the randomness of model selection for detection purposes. Recently, in the high-dimensional setting, \cite{bala2019} proposed and analyzed a diagnostic measure, known as the dif{}ference in model selection (DF(LASSO)), to directly quantify the influence of a single point on variable selection via LASSO. In addition, \cite{zhao2013} studied the so-called high-dimensional influence measure (HIM), capturing the influence of a single observation on the marginal correlation between the response and predictors. The versatility of this measure is then verified on variable selection via simulated and real data studies. This work is extended in \cite{zhao2015} and \cite{zhao2019} to detect multiple influential points, leading to the so-called multiple influential point (MIP) detection procedure. 


Yet, the aforementioned literature fall short on several fronts. First, are issues related to detecting multiple influential points. Indeed, multiple outliers are frequently encountered in fMRI-based studies \citep{mejia2017}. Their detection is challenging because of ``masking'' and ``swamping'' ef{}fects \citep{hadi1993}. Masking occurs when an influential point is hidden due to the presence of other influential cases, and swamping occurs when a non-influential observation is falsely labeled as influential. Existing methods are based on the leave-one-out scheme, which is not designed to overcome these ef{}fects when detecting multiple outliers. This is numerically corroborated in our simulation study (see Figures~\ref{fig:I_Y5_X0_SC8_n100_p1000} and \ref{fig:II_Y0_X10_SC8_n100_p1000}), as both DF(LASSO) and MIP underperform in capturing multiple influential points. In fact, in simulations resembling the aforementioned thermal pain prediction study, less than $10\%$ of the outliers are recognized by the two existing methods. Due to such inadequate detection,  heterogeneity is preserved in the data, leading to high prediction MSE even after outlier removal. Second, a gap exists in the theoretical framework of the DF(LASSO). Moreover, its extension to general penalty functions has not been explored, and the resulting statistical properties have not been carefully established. 

This manuscript aims to fill the aforementioned gaps and contribute to influential diagnosis on variable selection in five ways. First, upon characterizing influential points in assessing variable selection, we introduce the notion of sensitivity of model selectors, and draw connections with statistical concepts such as breakdown points and selection consistency. Second, we extend the diagnostic measure of \cite{bala2019} by bridging a missing step in its original theoretical development. Third, a generalized version accommodating universal model selectors is proposed and the corresponding asymptotic distribution determined. The main building blocks in establishing the large sample distribution of our proposed method are the preservation of exchangeability \citep{commenges2003} and characterization of a sequence of arbitrarily dependent events by its exchangeable counterpart \citep{galambos1973}. Fourth,  we propose a novel detection procedure, known as clustering-based MIP (ClusMIP), capable of identifying multiple influential observations. Fifth, we compare our proposal with competing procedures in comprehensive simulation and real data studies. ClusMIP is consistently shown to demonstrate significantly better detection performance than DF(LASSO). Its application to the scaled LASSO exhibits the highest detection outcome (Panels (A) of Figures~\ref{fig:I_Y5_X0_SC8_n100_p1000} and \ref{fig:II_Y0_X10_SC8_n100_p1000}). This implies that the (scaled) LASSO is the most sensitive model selector to influential points with regards to variable selection, and therefore serves as an appropriate candidate to use together with the ClusMIP procedure for outlier identification purposes. As such, our simulations of{}fers insights for practitioners into the sensitivity of dif{}ferent model selectors to outliers on variable selection performance under contamination. In addition, substantial enhancement in prediction and variable selection is observed by removing outliers detected by the ClusMIP. On the other hand, influential points detected by the ClusMIP (applied to MCP) for the fMRI-based pain prediction dataset strengthens the distinction between low and intense pain processing mechanisms. Furthermore, the resulting statistical model is scientifically interpretable and displays the best prediction performance after outlier removal. 

The manuscript is organized as follows: Section~\ref{sec:real_background} describes the motivating fMRI dataset \citep{lindquist2017} used to create predictive model for physical pain. Section~\ref{sec:principles} characterizes and defines influential points on variable selection.  Section~\ref{sec:method} begins with a survey of the existing HIM and DF(LASSO) approaches for single influential point detection. We then present our generalized version called the generalized dif{}ference in model selection (GDF) measure accompanied by the study of its asymptotic properties. We next discuss the MIP and propose our clustering-based detection scheme known as the ClusMIP to identify multiple influential observations. Section~\ref{sec:simulation} contains an intensive simulation study, followed by analysis of the fMRI dataset in Section~\ref{sec:realdata}. Conclusions and further discussions are given in Section~\ref{sec:conclusion}.  

\color{black}
\section{An fMRI Study of Thermal Pain}
\label{sec:real_background}

This dataset seeks to predict thermal-stimulated pain levels using multivariate patterns of brain activity and connectivity measured via fMRI data \citep{wager2013, lindquist2017}. It includes thirty-three healthy, right-handed participants with an average age of 27.9 years. Thermal stimulation was repeatedly delivered to the left volar forearm at six discrete temperatures: 44.3, 45.3, 46.3, 47.3, 48.3 and 49.3 $\tccentigrade$. After each stimulus, participants rated their pain on a 200-point scale. During the experiment, whole-brain fMRI data was acquired on an Achieva 3.0T TX MRI Scanner from Philips. Functional echo-planar imaging (EPI) images were acquired with: (\emph{i}) repetition time (TR)  = 2000 milliseconds (ms), (\emph{ii}) time to echo (TE) = 20 ms, (\emph{iii}) field of view (FOV) = 224 millimeters (mm), (\emph{iv}) matrix = 64 $\times$ 64, (\emph{v}) voxel size = 3.0 $\times$ 3.0 $\times$ 3.0 $\text{mm}^{3}$, and (\emph{vi}) acquisition parameters are 42 interleaved slices for parallel imaging with sensitivity encoding (SENSE) factor 1.5. Structural images were acquired using high-resolution T1 spoiled gradient recall images (SPGR) for anatomical localization and warping to a standard space. 

The structural T1-weighted images were co-registered to the mean functional image for each participant using the iterative mutual information-based algorithm implemented in the statistical parametric mapping (SPM) software package \citep{ashburner2005}, and were then normalized to the Montreal Neurological Institute (MNI) space. In each functional dataset, initial volumes were removed to allow for stabilization of image intensity. Prior to processing functional images, we removed volumes with signal values that were outliers within the time series. This was done by first computing the mean and the standard deviation of intensity values across each slice, for each image, and then computing the Mahalanobis distances for the matrix of slice-wise mean and standard deviation values. Values with a significant $\chi^2$-value (corrected for multiple comparisons based on the more stringent of either false discovery rate or Bonferroni methods) were considered outliers. Next, functional images were slice-time corrected and motion-corrected using SPM. Functional images were warped to SPM's normative atlas (warping parameters estimated from co-registered, high-resolution structural images) and interpolated to 2 $\times$ 2 $\times$ 2 $\text{mm}^{3}$ voxels.

The processed data was then parcellated into 489 regions using the Clinical Af{}fective Neuroimaging Laboratory (CANlab) combined whole-brain atlas. The dataset was further averaged across repeated trials at each of the six temperature levels, leading to a 198 by 489 matrix of brain activation. The pain ratings were similarly averaged over trials. 

Previous work on data using a similar experimental paradigm have shown  significant individual differences in how participants process pain that influence how different brain regions activate in response to a painful stimulus. Moreover, there is considerable inter-subject variability in functional topology in pain response, causing brain locations to be misaligned across participants \citep{wang2023improved}. We anticipate these issues could lead to substantial issues in variable selection and model performance if outlying participants are not properly accounted for. 
\color{black}

\section{Assessing Inf\hspace{+0.05em}luence on Variable Selection}
\label{sec:principles}

We now characterize and define the influential observations on variable selection. It is critical to first recognize that observations influencing the choice of a submodel can only be legitimately understood within a stochastic framework due to the data-dependent nature of model selection. Moreover, due to distinctive model selection criteria, especially dif{}ferent penalty functions of regularized variable selection methods, contrasting sets of predictors could be chosen by dif{}ferent model selectors. Thus, it is reasonable to conjecture that the extent to which dif{}ferent model selectors are perturbed by influential points in terms of selection consistency properties varies. An implication is that influential points in our context are specific to model selectors, and the same detection procedure may well lead to the detection of distinctive sets of influential points. 

In fact, the degree to which a model selector is prone to contamination can be termed as its {\em functional sensitivity} to influential observations. This concept can be understood as the percentage of data corruption permitted while fulfilling consistent selection properties. The significance of such sensitivity depends on the dichotomous intents of statistical operations, namely the diagnosis for contaminated data and consistent model selection. For diagnostic purposes, a model selector that is more sensitive to influential points may better facilitate their detection. On the other hand, the later objective of selection consistency underscores the resistance to contamination, which is closely linked to the notion of robustness of an estimator. Under this scenario, functional sensitivity philosophically aligns with the so-called breakdown point \citep{hampel1971, donoho1982} in robust statistics, which is the fraction of data allowed to be perturbed without invalidating the estimation consistency. From the above discussion, influential point set on variable selection is defined in a point-wise manner as:  

\begin{definition}{\textbf{Influential Observations on Variable Selection.}} \label{def:infl}
Observations indexed by $\text{I}_{\text{infl}} \subseteq \{1,...,n\} \in \mathcal{P} (\{1,...,n\})$ are called influential observations on variable selection if and only if for each $i \in \text{I}_{\text{infl}}$, $\widehat{\textbf{M}}_n \big(\textbf{Z}^{\big(\{i\} \cup \text{I}_{\text{infl}}^\mathsf{c}\big)} \big) \neq \widehat{\textbf{M}}_n\big(\textbf{Z}^{\big(\text{I}_{\text{infl}}^\mathsf{c}\big)}\big)$, where $\widehat{\textbf{M}}_n$ is model selector, which is a measurable function from the data to the space of all possible full-ranked submodels, $\textbf{Z}^{S}$ denotes the data with rows indexed by the set $S$ and $\text{I}_{\text{infl}}^\mathsf{c}=\{1,...,n\} \setminus \text{I}_{\text{infl}}$. 
\end{definition}

\section{Methodology}
\label{sec:method}

Let $\text{Y} \in \mathbb{R}$ be the response variable and ${x}_{j} \in \mathbb{R}, \, j=1,...,p$, be the predictors. Let $\textbf{Y} = (\text{Y}_{1},...,\text{Y}_{n})^{\top} \in \mathbb{R}^{n}$ be the response vector and $\textbf{X} = (\textbf{X}_{1},...,\textbf{X}_{n})^{\top} \in \mathbb{R}^{n \times p}$, where $\textbf{X}_{i}=(\text{X}_{i1},...,\text{X}_{ip})^{\top} \in \mathbb{R}^{p}$, $i=1,...,n$, be the $n \times p$ design matrix. Then, the $i^\text{th}$ observation or point of dataset $\textbf{Z} \coloneqq (\textbf{Y}, \textbf{X})$ is $\text{Z}_i \coloneqq (\text{Y}_{i}, \textbf{X}_{i}^\top)^\top \in \mathbb{R}^{p+1}$, $i=1,..,n$. Moreover, we abbreviate an arbitrary sequence of random variables or real values $a_1,...,a_n$ by $a_{[n]}$. 

A linear regression model with \emph{i.i.d.}\ errors is defined as
\begin{equation}
\text{Y}_{i}=  \textbf{X}_{i}^\top \boldsymbol{\beta} + \epsilon_i, \ i=1,...,n, \label{intro:linearmodel}
\end{equation} 
where $\boldsymbol{\beta}=(\beta_1,...,\beta_p)^{\top} \in \mathbb{R}^{p}$ is the true regression coef{}ficient, and $\epsilon_{i} \sim N(0,\sigma^2)$ for some $\sigma^2 > 0$. Moreover, the $\epsilon_{i}$'s are independent of $\textbf{X}$. 

\subsection{Single Inf\hspace{+0.05em}luential Point Detection}
\label{sec:method_single}


The HIM and DF(LASSO) are revised in Sections~\ref{sec:method_him} and \ref{sec:method_bala_single}, followed by proposing a generalized version in Section~\ref{sec:method_modified_single}.


\subsubsection{High-dimensional Inf\hspace{+0.05em}luence Measure (HIM)}
\label{sec:method_him}
Based on the linear regression model \eqref{intro:linearmodel}, \cite{zhao2013} proposed a non-parametric, model-free and theoretically justified diagnostic measure capable of {\em indirectly} assessing the influence of a single observation on variable selection for high-dimensional data. Known as the high-dimensional influence measure (HIM), this metric is formulated on the leave-one-out basis and is designed to quantify the contribution of each individual observation to the marginal correlation between the response variable and all the predictors. An observation with excessively large marginal correlation is deemed influential. To compute the HIM, we first calibrate the influence of the $i^\text{th}$ observation on the marginal correlation between the response variable $\text{Y}$ and $j^\text{th}$ predictor $x_j$ by calculating its sample estimate $\widehat{\rho}_j^{(i)} = \sum_{k=1, k \neq i}^{n} \big \{(\text{X}_{kj}-\widehat{\mu}_{xj}^{(i)})(\text{Y}_k-\widehat{\mu}_y^{(i)}) \big \}/ \big \{(n-1)\widehat{\sigma}_{xj}^{(i)} \widehat{\sigma}_y^{(i)} \big \}$, $j=1,...,p$ and $i=1,...,n$,  where $\widehat{\mu}_{xj}^{(i)} = \sum_{k=1, k \neq i}^{n} \text{X}_{kj}/n$, $\widehat{\mu}_y^{(i)} = \sum_{k=1, k \neq i}^{n} \text{Y}_k/n$, $\big(\widehat{\sigma}_{xj}^{(i)}\big)^2 = \sum_{k=1, k \neq i}^{n}(\text{X}_{kj}-\widehat{\mu}_{xj}^{(i)})^2/(n-1)$, and $\big(\widehat{\sigma}_y^{(i)}\big)^2=\sum_{k=1,k \neq i}^{n}(\text{Y}_k-\widehat{\mu}_y^{(i)})^2/(n-1)$. Then, upon aggregating and averaging over all $p$ predictors, the HIM for the $i^\text{th}$ observation is given by $D_i = \sum_{j=1}^{p} \big(\widehat{\rho}_j-\widehat{\rho}_j^{(i)}\big)^2/p, \ i=1,...,n$, where $\widehat{\rho}_j^{(i)}$ is given above and $\widehat{\rho}_j$ is its counterpart computed based on the full dataset. Under certain conditions (\emph{ibid.}, Conditions C.1 - C.4, Section 2.3), \cite{zhao2013} establish that in the absence of influential points, $n^2 D_i \rightarrow \chi^2(1) \ \text{as} \ \text{min}(n,p) \rightarrow \infty, \ i=1,...,n$, where $\chi^2(1)$ is the chi-squares distribution with one degree of freedom. Thus, the scaled percentile of $\chi^2(1)$ can be used as a threshold to pinpoint influential observations. The authors then demonstrate the potential to extend their framework to GLM, yet remarking that the above asymptotic distributional property may not hold. The ef{}fectiveness of this method in assessing influence on variable selection is further validated via simulated studies.

Despite advantages including computational simplicity and sound theoretical justification, there are limitations of the HIM approach. First, since it is primarily proposed under the linear regression model, its ef{}fectiveness when the response variable is nonlinearly correlated with the predictors is ambiguous. Second, when $p$ is close to $n$, spurious correlation may exist between the response variable and predictors leading to unreliable variable selection, an undesirable phenomenon known as the Freedman's paradox \citep{freedman1983} which persists in high-dimensional settings \citep{fan2014}. Thus, adopting marginal correlation to evaluate influence on variable selection may be inappropriate. Third, the HIM approach is independent of model selector, thus providing no insights of the sensitivity of a model selector in spite of the fact that it is not intended to perform such function. 

\subsubsection{Dif{}ference in Model Selection via the LASSO (DF(LASSO))}
\label{sec:method_bala_single}
In comparison with \cite{zhao2013}, under \eqref{intro:linearmodel}, \cite{bala2019} formulated the so-called dif{}ference in model selection (DF(LASSO)) measure to {\em directly} gauge the influence of a single point on the submodel selected by the LASSO. Similar to the HIM, the DF(LASSO) is also based on the leave-one-out scheme and is defined by $\iota_i = (\delta_i-\mathbb{E}(\delta_i))/\sqrt{\Var(\delta_i)}$, $i=1,...,n$, where  
\begin{equation}
\delta_i = \sum_{j=1}^{p} \big \lvert \mathbbm{1}(\widehat{\beta}_{\text{LASSO},j}=0) - \mathbbm{1}( \widehat{\beta}^{(i)}_{\text{LASSO},j}=0) \big \rvert. \label{balameasure}
\end{equation}
Here, $\mathbbm{1}(\cdot)$ is the indicator function, $\widehat{\beta}_{\text{LASSO},j}$ and $\widehat{\beta}^{(i)}_{\text{LASSO},j}$ respectively denote the $j^\text{th}$ component of the LASSO estimates obtained from the full dataset and reduced dataset with $i^\text{th}$ observation removed. Then, assuming $\sum_{j=1}^{p} \mathbbm{1} (\beta_j=0) \rightarrow \infty$, the authors verify that $(\Delta-\mathbb{E}(\Delta))/\sqrt{\Var(\Delta)} \rightarrow N(0,1) \ \text{as} \ \text{min}(n,p) \rightarrow \infty$, where $\Delta = \sum_{j=1}^{p} \big \lvert \mathbbm{1}(\widehat{\beta}_{\text{LASSO},j}=0) - \mathbbm{1}(\beta_{j}=0) \big \rvert$ and $N(0,1)$ is a standard Gaussian. 

In practice, the sample mean and variance of the $\delta_i$'s in \eqref{balameasure} are suggested to estimate $\mathbb{E}(\delta_i)$ and $\Var(\delta_i)$ in the DF(LASSO) measure. For diagnosis, an observation $i$ with $\lvert \iota_i \rvert \geq 2$ is considered influential by the asymptotic Gaussian approximation. From the formulation of the $\delta_i$'s, we see that they play central roles in the assessment as they directly quantify the fluctuation in model selection. Moreover, the comparative magnitude of each $\delta_i$ in relation to others, rather than its absolute scale, is more important for decision making. In other words, $\delta_i$'s shall be viewed from a collective angle to ascertain the nature of any individual observation. Indeed, this interpretation underpins an important perspective that every individual point is influential on variable selection, yet some may be more influential than the others. 

Comparing the DF(LASSO) with the HIM, we observe that in terms of methdological philosophies, while the DF(LASSO) explicitly captures the deviation in the selected submodels via changes in the sparsity of the LASSO regression coef{}ficients obtained from the full and reduced datasets, the HIM takes advantage of the marginal correlation as the primary diagnosis instrument, leading to influence induced in an implicit, latent manner. On the computational front, the DF(LASSO) requires fitting $n+1$ models, while no such process is required for the HIM. Thus, the HIM is more computational ef{}ficient. 

Apart from the above characteristics, there are limitations of the DF(LASSO) framework. First, in establishing the asymptotic normality, the authors recommend using the sample mean and variance of $\delta_{[n]}$ to estimate $\mathbb{E}(\delta_i)$ and $\Var(\delta_i)$, which is not theoretically substantiated. Second, such asymptotic normality is specifically derived for the LASSO estimator. The applicability of this theoretical framework to other model selectors is yet to be established. 

\subsubsection{Generalized Dif{}ference in Model Selection Measure (GDF)}
\label{sec:method_modified_single}
We now develop a universally feasible diagnostic measure by addressing two limitations of the DF(LASSO) method, namely: (\emph{i}) the legitimacy of estimating $\mathbb{E}(\delta_i)$ and $\Var(\delta_i)$ by the sample mean and variance, and (\emph{ii}) applicability to extend the framework to other model selectors. To achieve that, we note that $\delta_{[n]}$ is a sequence of dependent random variables and each $\delta_i$ is a sum of correlated Bernoulli random variables, where both dependence have no tractable forms. 

To address the first issue regarding the eligibility of estimating $\mathbb{E}(\delta_i)$ and $\Var(\delta_i)$ by the corresponding sample mean and variance, we first focus on the finite-sample property of $\delta_{[n]}$ and bridge the gap between this sequence and the characteristic of the original dataset. With Theorem~\ref{thm:delta_identical} reported below, we show that $\delta_{[n]}$ is exchangeable and thus identically distributed. The proof of Theorem 1 is given in Section 2 of the Supplementary Materials Part I (Zhang, Asgharian and Lindquist (2026)). 

\newtheorem{thm}{Theorem}
\begin{thm}[]\label{thm:delta_identical}
Suppose that the dataset $\textbf{Z}=(\text{Z}_1,...,\text{Z}_n)^\top$, where $\text{Z}_i = (\text{Y}_i, \textbf{X}_{i}^\top)^\top \in \mathbb{R}^{p+1}, \ i=1,...n$, has \emph{i.i.d.}\ rows. Then, the sequence of non-negative, discrete diagnostic measures $\delta_{[n]}$ defined in \eqref{balameasure} is exchangeable and thus marginally identically distributed. 
\end{thm}

From Theorem~\ref{thm:delta_identical}, we next provide two relevant results of an exchangeable sequence of random variables in Corollary 1, which are the consistency of the empirical cumulative distribution function (ECDF) and strong law of large numbers (SLLN). The combined statements further strengthen the theoretical basis for estimating $\mathbb{E}(\delta_i)$ and $\Var(\delta_i)$ by the sample mean and variance. Its proof is given in Section 3 of the Supplementary Materials Part I (Zhang, Asgharian and Lindquist (2026)). 

\begin{corollary}[]\label{corollary:convergence}
Let $\delta_{[n]}$ in \eqref{balameasure} be defined on a probability space $(\Omega, \mathcal{F}, \mathbb{P})$, and let $\mathcal{F}_{\infty}$ be its tail $\sigma$-algebra. Moreover, let $\bar{\delta} \coloneqq \frac{1}{n} \sum_{i=1}^{n} \delta_i$. Then, as $n \rightarrow \infty$, we have 
\begin{enumerate}
    \item Uniform Convergence of ECDF: $\sup_{x \in \mathbb{R}} \vert \widehat{F}_n - F \vert \xrightarrow{\text{a.s.}} 0$, where $\widehat{F}_n(x) = \sum_{i=1}^{n} \mathbbm{1}(\delta_i \leq x)$, $x \in \mathbb{R}$ and $F$ is the true cumulative distribution function (CDF) of $\delta_{[n]}$. In particular, the discontinuity of the limiting CDF $F$ is permitted. 

    \item SLLN: $\frac{1}{n} \sum_{i=1}^{n} \delta_i \xrightarrow{\text{a.s.}} \mathbb{E}(\delta_i \vert \mathcal{F}_{\infty})$ and $\frac{1}{n} \sum_{i=1}^{n} (\delta_i-\bar{\delta})^2 \xrightarrow{\text{a.s.}} \Var(\delta_i \vert \mathcal{F}_{\infty})$.    
\end{enumerate}

\end{corollary}

To address the second issue on the applicability of the DF(LASSO) framework to other model selectors, it is crucial to recognize that the crux of the problem indeed lies in ascertaining the finite- or large-sample distribution of $\tau_{[n]}$, which is the counterpart of $\delta_{[n]}$ in \eqref{balameasure} without model selector restriction and is therefore component-wise defined by 
\begin{equation}
\tau_i = \sum_{j=1}^{p} \xi_{i,j} \ \text{and} \ \xi_{i,j} \coloneqq \lvert \mathbbm{1}(\widehat{\beta}_{j}=0) - \mathbbm{1}(\widehat{\beta}^{(i)}_{j}=0) \rvert, i=1,...,n, \label{gdf}
\end{equation}
where $\widehat{\beta}_{j}$ and $\widehat{\beta}^{(i)}_{j}$ are the $j^\text{th}$ components of a generic sparse regression coef{}ficient estimate obtained based on the full and reduced data with the $i^\text{th}$ observation removed, respectively. As such, $\tau_i$ in \eqref{gdf} is referred to as the generalized dif{}ference in model selection (GDF) measure in the sequel. To understand the distributional properties of $\tau_i$, both parametric and non-parametric approaches can provide reliable and theoretically appropriate approximations. In this article, we specifically focus on one such avenue by theoretically justifying the central limit theorem (CLT) for each $\tau_i$ via an approach distinctive from \cite{bala2019}. 

To verify the CLT for $\tau_i$, we bridge the theoretical hiatus in \cite{bala2019} for generic model selectors through the work of \cite{galambos1973}, which bypasses the challenge in handling the intractable dependence among its summands $\xi_{i,j}$. Specifically, it is shown [\emph{ibid.}, Theorem 1] that for an arbitrary sequence of events, there always exists a sequence of exchangeable events such that the respective sums are equally distributed in finite sample. This implies that $\tau_i$ in \eqref{gdf}, the sum of dependent indicators $\xi_{i,j}, j=1,...,p$, is identically distributed as the sum of a sequence of exchangeable Bernoulli random variables. Toward that end, the regular \citep{klass1987} and empirical \citep{hahn1998} CLT for an exchangeable sequence of random variables collectively facilitate the derivation of the following assertion, where its proof is available in Section 4 of the Supplementary Materials Part I (Zhang, Asgharian and Lindquist (2026)): 

\begin{thm}[]\label{thm:tau_clt}
Let $\tau_i$ be the GDF measure defined as in \eqref{gdf}. Suppose that the conditions (A.1) and (A.2) (see Section 4 of the Supplementary Materials Part I (Zhang, Asgharian and Lindquist (2026))) hold. Then, 
\begin{equation}
\frac{\tau_i-\bar{\tau}}{\sigma_\tau} \rightarrow N(0,1) \ \text{as} \ \text{min}(n,p) \rightarrow \infty, \label{gdf_clt} 
\end{equation}
where $\bar{\tau}$ and $\sigma^2_\tau$ are the sample mean and variance of the sequence $\tau_{[n]}$. 
\end{thm}

While we essentially have summation of Bernoulli variables, we have presented the conditions in rather general form to extend the applicability of the result. Here, condition (A.1) controls the tail behavior of the exchangeable summand of $\tau_i$ by requiring that it is not heavy tailed, which is a weak version of the existence of the first moment. Condition (A.2) is parallel to the classical Lindeberg's condition in establishing the CLT, which requires that the contribution of each individual summand to the variance of $\tau_i$ is suf{}ficiently small. In fact, the existence of a uniform bound on the second moment of the summand serves as a sufficient (not necessary) condition for (A.1) and (A.2). In our case, each individual summand is Bernoulli so that both conditions are fulfilled. From \eqref{gdf_clt}, at a given nominal level $\alpha \in (0,1)$, the $100 \cdot (1-\alpha/2)^\text{th}$ percentile of $N(0,1)$ is a valid threshold such that a point with an absolute scale of the standardized $\tau_i$ exceeding that cut-of{}f is deemed influential. Moreover, to test the two-sided hypothesis that the $i^\text{th}$ point is clean, the $p$-value is $2 \cdot \mathbb{P}(N(0,1) > (\tau_i-\bar{\tau})/\sigma_\tau)$. A framework for detecting a single influential point via the GDF metric is given in Algorithm GDF in Section 5 of the Supplementary Materials Part I (Zhang, Asgharian and Lindquist (2026)). 

\subsection{Multiple Inf\hspace{+0.05em}luential Observations Detection}
\label{sec:method_multiple}
To detect multiple influential observations, we first review the multiple influential point (MIP) detection algorithm by \cite{zhao2019} in Section~\ref{sec:mip}, which is an extension of the HIM procedure discussed in Section~\ref{sec:method_him}. This is followed by the proposal of our clustering-based multiple influential point (ClusMIP) detection method in Section~\ref{sec:clusmip}. 

\subsubsection{Multiple Inf\hspace{+0.05em}luential Point (MIP) Detection Procedure}
\label{sec:mip}

The leave-one-out basis of the HIM approach implies that it may not be ef{}fective in capturing multiple influential points. To remedy this issue, \cite{zhao2019} propose the MIP method via integrating the so-called random-group-deletion (RGD) scheme into the HIM framework.  Based on this, two HIM-based statistics are subsequently formulated to overcome both masking and swamping ef{}fects in detecting multiple influential observations. 

To have a proper understanding of this procedure, we first explain the philosophy underlying the RGD algorithm, which is provided in Section 6 of the Supplementary Materials Part I (Zhang, Asgharian and Lindquist (2026)). The RDG scheme centers on seeking and  subsequently attaching a clean subset of data to every single point, of which refinement is conducted based on the combined dataset. To be more precise, for each observation $i$, $m$ samples indexed by $A_1,...,A_m$ are drawn uniformly and randomly from the remaining portion of the data. When $m$ is suf{}ficiently large, there is a high probability that the subset $\{\text{Z}_r: r \in A_j\}$ is clean for some $j \in \{1,...,m\}$. Thus, if $i^\text{th}$ observation is contaminated, it would be the sole influential point in the merged subset $\{\text{Z}_r: r \in A_j \cup \{i\}\}$. Thus, further assessment can be carried out on this combined dataset. 

Toward that end, for each $i$, two HIM-based statistics $\text{T}_{\text{min},i,m}$ and $\text{T}_{\text{max},i,m}$ are then  constructed, which are respectively defined by $\text{T}_{\text{min},i,m} = \argmin_{1 \leq j \leq m} n_\text{sub}^2 D_{j,i}$ and $\text{T}_{\text{max},i,m} = \argmax_{1 \leq j \leq m} n_\text{sub}^2 D_{j,i}$, where $D_{j,i} = \sum_{k=1}^{p} \big(\widehat{\rho}_k^{(A_j \cup \{i\})}-\widehat{\rho}_k^{(A_j}\big)^2/p$. Here, $\widehat{\rho}_k^{(A_j \cup \{i\})}$ and $\widehat{\rho}_k^{(A_j)}$ are the sample marginal correlations between the response $\text{Y}$ and $k^\text{th}$ predictor $x_k$ based on $\{\text{Z}_r: r \in \{A_j \cup \{i\}\}\}$ and $\{\text{Z}_r: r \in A_j\}$. The authors then establish that they are both asymptotically $\chi^2(1)$ and indeed complement each other: $\text{T}_{\text{min},i,m}$ tends to select conspicuously influential points. Thus, it is ef{}fective in overcoming the swamping ef{}fect. However, it may be conservative so that true influential points may be neglected. In contrast, $\text{T}_{\text{max},i,m}$ is powerful to counter the masking ef{}fect. Yet, it may be inordinately aggressive so that clean ones may be falsely chosen. Therefore, combining the strengths of both $\text{T}_{\text{min},i,m}$ and $\text{T}_{\text{max},i,m}$ leads to the MIP detection procedure provided in Section 7 of the Supplementary Materials Part I (Zhang, Asgharian and Lindquist (2026)). In practice, the MIP algorithm is implemented in the $\tt{R}$ package $\tt{MIP}$.  

\subsubsection{Clustering-based MIP (ClusMIP) Detection Procedure}
\label{sec:clusmip}

Motivated by \cite{zhao2015, zhao2019}, one apparent solution to detect multiple influential observations on variable selection is to integrate the RGD scheme into the GDF metric \eqref{gdf}, leading to the RGD-based GDF (RGDF) procedure provided in Section 8 of the Supplementary Materials Part I (Zhang, Asgharian and Lindquist (2026)).

Indeed, examination of both the MIP and RGDF procedures shows that they consist of two broad stages, the preliminary filtering and secondary refining phases. The main purpose of the filtering stage is to obtain an estimate of the potentially influential points. Then, this set is refined to remove false selections, leading to the final estimate of influential points. In fact, this agrees with our perspective of influential points on variable selection provided in Definition~\ref{def:infl}, where the complete dataset is separated into clean and contaminated parts so that the concept of an influential point can be unambiguously understood and formalized. 

On the other hand, it is crucial to note that the detection outcomes of the MIP and RGDF procedures largely hinge on the accuracy (ef{}fectiveness) and the accompanying computational cost (ef{}ficiency) of the filtering stage in producing a reliable set of influential points. The MIP can be deemed both ef{}fective and ef{}ficient since the computations of $\text{T}_{\text{min},i,m}$ and $\text{T}_{\text{max},i,m}$ are model-free. Yet, in our case, for a fixed number of random draws $m$ with $n_\text{size}$ being the cardinality of each random sample, a maximum of $nm(n_\text{size}+2)$ final fitted models are obtained. This is computationally expensive even if $n$, $m$ and $n_\text{size}$ are moderate. Thus, ef{}ficiency could be severely compromised in the RGDF framework when iterative model fitting is involved to assess influence on variable selection. 


To alleviate the computational costs, instead of the RGD strategy, we incorporate an appropriate high-dimensional clustering procedure for initially partitioning the complete dataset into approximately clean ($\widehat{\text{S}}_{\text{clean}}$) and influential ($\widehat{\text{S}}_{\text{infl}}$) portions. Then, each point $k$ in $\widehat{\text{S}}_{\text{infl}}$ is further assessed solely based on the subset of the data indexed by $\{k\} \cup \widehat{\text{S}}_{\text{clean}}$ via Algorithm GDF which is proposed for detecting a single influential point through the GDF measure \eqref{gdf}. Moreover, to simultaneously test the $\rvert \widehat{\text{S}}_{\text{infl}} \lvert$ hypotheses that the $k^\text{th}$ point is clean, $k \in \widehat{\text{S}}_{\text{infl}}$, the procedure of \cite{benjamini1995} is adopted to address the multiple testing issue. This leads to our ClusMIP procedure summarized below in Algorithm~\ref{algo:clusmip}. 

\begin{varalgorithm}{ClusMIP} 
\caption{Clustering-based Multiple Influential Point Detection (ClusMIP)}
\label{algo:clusmip} 
\begin{algorithmic}[1]
\State Partition the dataset into the potentially influential ($\widehat{\text{S}}_{\text{infl}}$) and clean ($\widehat{\text{S}}_{\text{clean}}$) proportions via an appropriate high-dimensional clustering procedure, where $\lvert \widehat{\text{S}}_{\text{infl}} \rvert < \vert \widehat{\text{S}}_{\text{clean}} \rvert$.  

\State Given a nominal level $\alpha \in (0,1)$, the final estimated influential observations set is $\widehat{\text{I}}_{\text{infl}} = \{j \in \widehat{\text{S}}_{\text{infl}}: \lvert (\tau_j-\tilde{\tau}_j)/\tilde{\sigma}_{\tau_j} \rvert > \gamma_{\alpha} \}$, where for each $j \in \widehat{\text{S}}_{\text{infl}}$, $\tau_j$ defined in \eqref{gdf} is computed from subset indexed by $\{j\} \cup \widehat{\text{S}}_{\text{clean}}$, and $\tilde{\tau}_j$ and $\tilde{\sigma}_{\tau_j}$ are the sample mean and variance of $\{\tau_k: k \in \{j\} \cup \widehat{\text{S}}_{\text{clean}} \}$. Here, $\gamma_{\alpha}$ is the $100\cdot(1-\alpha/2)^\text{th}$ percentile of a standard normal $N(0,1)$. To simultaneously test hypotheses that $k^\text{th}$ point is not influential, $k \in \widehat{\text{S}}_{\text{infl}}$, procedure of \cite{benjamini1995} is adopted.   
\end{algorithmic}
\end{varalgorithm}   

The intent and accompanying advantage of integrating a clustering operation is four-fold. First, it mitigates the computational requirement via of{}fering a fast and conveniently implementable solution. Second, the heterogeneity ingrained in the data, which contributes to the discrepancy in variable selection, could be captured by the resulting partition outcomes. Third, despite being potentially inaccurate, such partition outcomes are then refined in the second stage via further assessment through the GDF measure, safeguarding the detection accuracy. Lastly, as shown in Theorem~\ref{thm:clusmip_fpr} and verified in the simulated study (Section~\ref{sec:simulation}), the integration of such clustering controls the false positive rate (FPR) at a given false discovery rate (FDR) $\alpha_0 \in (0,1)$ in any finite sample. Indeed, the FPR diminishes as $\min(n,p) \rightarrow \infty$ if the clustering is strongly consistent in estimating the true cluster centers. Thus, it remains to pinpoint a reasonable high-dimensional clustering candidate.

Toward that end, one possible approach is $K$-means clustering (KC), which minimizes the within-cluster Euclidean distance. \cite{pollard1981} provides conditions for the strong consistency of the KC, showing that with the correct number of clusters, the cluster centers converge to their true counterparts as $n \rightarrow \infty$ under certain conditions. Yet, this approach is limited on three facets. First, its widely adopted approximating algorithm, known as Lloyd's algorithm, may not be globally optimal due to its heavy sensitivity to the starting points. Second, the consistency of the KC may not hold under diverging dimension $p \rightarrow \infty$. Third, under such high-dimensional settings, the notion of distance in terms of the $\ell^p$ norm may not be meaningful \citep{beyer1999}. Indeed, it is justified [\emph{ibid.}, Theorem 1] that as $p \rightarrow \infty$, the ratio of the distances in $\ell^p$ norm of the nearest and furthest points to a reference point approaches 1, implying that points tend to be uniformly separated in the high-dimensional $\ell^p$ space. 

To address the first limitation, an improved choice of the initial values for the KC algorithm is proposed by \cite{arthur2007}, leading to the so-called $K$-means++ algorithm (KC++). In addition, to resolve the challenges brought about by the high dimensionality, two categories of remedial proposals are available in the literature, which are known as the projection-based and regularized clustering. The projection-based scheme is a hybrid technique by first projecting the high-dimensional data onto a low-dimensional space, followed by partition via the KC. Two proposals under this category are the $t$-distributed stochastic neighbor embedding ($t$-SNE) \citep{hinton2002} and spectral clustering \citep{ng2001}, where the projected low-dimensional space for the later work is spanned by eigenvectors of the normalized graph Laplacians. In comparison, the regularized clustering augments a penalty term to the appropriate objective function to obtain estimates of penalized cluster centers. One such approach is called regularized KC \citep{sunwang2012}, where a group LASSO penalty term is adopted and strong consistency of cluster centers is established [\emph{ibid.}, Theorem 1] for $p \rightarrow \infty$. These five clustering procedures, including the KC, KC++, $t$-SNE, spectral clustering and regularized KC are empirically compared on all the simulation configurations considered in Section~\ref{sec:simulation}. The corresponding graphical illustrations are reported in Section 10 of the Supplementary Materials Part I (Zhang, Asgharian and Lindquist (2026)), where all of them exhibit analogously satisfactory detection of clusters. Indeed, in both finite and diverging dimensions, with a consistent clustering technique that is further refined via the CLT of the GDF measure \eqref{gdf} established in Theorem~\ref{thm:tau_clt}, the FPR of the foregoing ClusMIP algorithm is controlled at a given the FDR which is stated below in Theorem~\ref{thm:clusmip_fpr}. The consistency conditions 
 (B.1) to (B.2) \citep{pollard1981}, (C.1) to (C.7) \citep{sun2012}, and the proof are given in Section 9 of the Supplementary Materials Part I (Zhang, Asgharian and Lindquist (2026)):

\begin{thm}[]\label{thm:clusmip_fpr}
Suppose that the proportion of clean observations is more than 1/2 and FDR is controlled at $\alpha_0 \in (0,1)$. Then, 
\begin{enumerate}
    \item[(a)] $\mathbb{E}[\text{FPR}(\text{ClusMIP})] \leq \alpha_0$ at any finite dimensions $n$ and $p$. 

    \item[(b)] If $p$ is fixed, then $\mathbb{E}[\text{FPR}(\text{ClusMIP})] \rightarrow 0$ as $n \rightarrow \infty$ provided that (B.1) and (B.2) hold.

    \item[(c)] If $p \rightarrow \infty$, then $\mathbb{E}[\text{FPR}(\text{ClusMIP})] \rightarrow 0$ as $n \rightarrow \infty$ provided that (C.1) to (C.7) hold.
\end{enumerate}
where the FPR(ClusMIP) refers to the FPR of the ClusMIP given in Algorithm~\ref{algo:clusmip}.
\end{thm}  

The aforementioned conditions (B.1) to (B.2) and (C.1) to (C.7) are standard assumptions in clustering. They ensure the uniqueness of the true cluster centers. Moreover, they require that the second moment of the distribution of data points is finite, and the distribution is continuous. Condition (C.5) is a technical condition requiring the existence of a spherically symmetric envelop for the underlying density whose $p^{\text{th}}$ moment exists. This assumption requires some control on the tail behavior of the underlying density. It essentially says that the tails of the density of data points should vanish polynomially, of degree $p$ fast. As a result, the higher the dimension of the space, $p$, the faster the tails should die out. 

\section{Simulation Study} 
\label{sec:simulation}

The data generation is explained in Section~\ref{sec:datageneration}, followed by evaluation metric and implementation in Sections~\ref{sec:assessment} and \ref{sec:methodscodesandimplementation}. Analysis of results is in Section~\ref{sec:simulationresults}. 

\subsection{Data Generation} The specifics of data generation are given as follows: 
\label{sec:datageneration}

\paragraph*{\textbf{Design Matrices $\textbf{X}$}} 
Let $(n,p)=(100,1000)$ and $(160, 480)$, where the latter dimension setting resembles those observed in the motivating fMRI study \citep{lindquist2017}. For each set of dimensions, $\textbf{X} \in \mathbb{R}^{n \times p}$ follows a zero-mean multivariate normal distribution, denoted $\textbf{X} \sim \mathcal{N}(\boldsymbol{0}, \Sigma)$. Two possible variance-covariance matrices $\Sigma \in \mathbb{R}^{p \times p}$ are considered: the Toeplitz and the spatially correlated structures. Specifically, their $(i,j)$-entry is given by $\rho_{1}^{|i-j|}$ and $\rho_{2}^{|i-j|} \mathbbm{1}(i,j \in \{(p-S)/2+1,...,(p-S)/2+S\})$, $i=1,...,p$ and $j=1,...,p$, respectively. The spatially correlated formulation, where only the center $S$ predictors are dependent, mimics the brain connectivity patterns observed in fMRI. In our simulation study, we let $\rho_1 \in \{0,0.5,0.8\}$, $\rho_2=0.8$ and $S=100$ so that four design matrices respectively denoted by $\textbf{X}_{\text{Tp(0)}}$, $\textbf{X}_{\text{Tp(0.5)}}$, $\textbf{X}_{\text{Tp(0.8)}}$ and $\textbf{X}_{\text{Sc(0.8)}}$ are generated for each dimension.

\paragraph*{\textbf{Influential and Clean Observations}} 
Influential and clean observations are synthesized using the following two-step procedure: first, generate $n=100$ observations based on \eqref{intro:linearmodel}. Second, generate $n_\text{infl}$ points, denoted by $\{\textbf{Z}_{i,\text{infl}} = (\text{Y}_{i,\text{infl}}, \textbf{X}_{i,\text{infl}}^\top)^\top \in \mathbb{R}^{1001}: i=1,...,n_\text{infl} \}$, according to three perturbation schemes presented below, where $n_\text{infl}$ is the number of influential points. The first $n_\text{infl}$ points in the first step are then replaced by $\textbf{Z}_{i,\text{infl}}, i=1,..,n_\text{infl}$, yielding the  heterogeneous, contaminated dataset. Here, the three schemes respectively introduce contamination to the response vector only, design matrix only and both of them simultaneously. Moreover, the proportion of contamination $\zeta \coloneqq (n_\text{infl}/n) \times 100\%$ is set to be $5\%, 10\%, 15\%$ and $20\%$. The data generation is given in Section 11.1 of the Supplementary Materials Part I (Zhang, Asgharian and Lindquist (2026)). 

\paragraph*{\textbf{Perturbation Schemes}} The true regression coef{}ficient in \eqref{intro:linearmodel} is $\boldsymbol{\beta}=\{2,1,1,0,...,0\}^{\top} \in \mathbb{R}^{p}$ and $\sigma^2=1$. To generate influential points, we define $i_{\text{linear},\text{max}}=\argmax_{1 \leq i \leq n} \text{Y}_i$ to facilitate the distinction of them from the clean portion based on the chosen clustering method. Then, for $i=1,...,n_\text{infl}$, we adapt and modify the settings in \cite{zhao2013} and propose the following three schemes each with low, moderate and high contamination intensities/scales:  
\begin{enumerate}

\item \text{Scheme I (Contamination on the Response)}: $\textbf{X}_{i,\text{infl}} = \textbf{X}_{i}$ and $\text{Y}_{i,\text{infl}} = \text{Y}_{i_{\text{linear},\text{max}}} \\ + \epsilon_i + \kappa_{\text{linear}, \textbf{Y}}$, where $\epsilon_i \sim N(0,1)$ and $\kappa_{\text{linear}, \textbf{Y}} \in \{5,10,20\}$. 

\item \text{Scheme II (Contamination on the Predictors)}: $\text{X}_{ij,\text{infl}} = \text{X}_{ij} + \kappa_{\text{linear}, \textbf{X}} \cdot \mathbbm{1}\big\{j \in \{1,...,10\} \big\}$, $\kappa_{\text{linear}, \textbf{X}} \in \{5,10,20\}$, $\textbf{X}_{i,\text{infl}} = (\text{X}_{i1,\text{infl}},...,\text{X}_{ip,\text{infl}})^{\top}$, and $\text{Y}_{i,\text{infl}} = \text{Y}_{i}$.  

\item \text{Scheme III (Contamination on the Response and Predictors)}: $\text{X}_{ij,\text{infl}} = \\ \text{X}_{i_{\text{linear},\text{max}}j} + 0.5 \ \kappa_{\text{linear}, \textbf{X}} \cdot \mathbbm{1}\big\{j \in \{1,...,10\} \big\}$, $\textbf{X}_{i,\text{infl}} = (\text{X}_{i1,\text{infl}},...,\text{X}_{ip,\text{infl}})^{\top}$, and $\text{Y}_{i,\text{infl}} = (\textbf{X}_{i,\text{infl}})^{\top} \boldsymbol{\beta} + \epsilon_i + \kappa_{\text{linear}, \textbf{Y}}$, where $\epsilon_i \sim N(0,1)$ and $(\kappa_{\text{linear}, \textbf{X}},\kappa_{\text{linear}, \textbf{Y}}) \in \{(5,5), (10,10),(20,20)\}$. 
\end{enumerate}

\subsection{Assessment Metrics}
\label{sec:assessment}
Two groups consisting of six metrics were computed The first group consists of the power ($\text{Power}_{\text{Detection}(\text{Selector})}$), false positive rate of detection ($\text{FPR}_{\text{Detection}(\text{Selector})}$), mean number of detected influential points impacting the selection of stable/true signals/predictors ($\text{Stable}_{\text{Detection}(\text{Selector})}$), and computation time ($\text{Time}_{\text{Detection}(\text{Selector})}$). Specifically, $\text{Stable}_{\text{Detection}(\text{Selector})}$ determines whether each individual identified outlier alters the choice of at least two (out of a total three) pre-specified data-generating predictors. Here ``Detection(Selector)'' specifies the identification procedure ``Detection'' when a submodel is selected by ``Selector''. They directly evaluate the detection performance. The second group includes the MSE of prediction and the empirical probability that the pre-specified true model is contained in the selected submodel (abbreviated as above), each computed both before and after removal of detected influential points for comparison purposes. They highlight the benefits of outlier detection on downstream inference. Here, since MIP has no model selection, the latter two metrics are excluded. These six metrics are formally defined in Section 11.2 of the Supplementary Materials Part I (Zhang, Asgharian and Lindquist (2026)). 

\subsection{Methods and Implementation}
\label{sec:methodscodesandimplementation}
Three detection methods are included: the ClusMIP, DF(LASSO) and MIP. In case of $\textbf{X}_{\text{Tp(0.5)}}$, $\textbf{X}_{\text{Tp(0.8)}}$ and $\textbf{X}_{\text{Sc(0.8)}}$, the ClusMIP is combined with LASSO, scaled LASSO (SLASSO), elastic net (ENET), SCAD and MCP, while the ENET is excluded for $\textbf{X}_{\text{Tp(0)}}$. Moreover, tuning parameter for all the regularized model selectors is chosen via the 10-fold cross-validation (CV), unless otherwise specified. A summary of methods and implementation is given in Section 11.3 of the Supplementary Materials Part I (Zhang, Asgharian and Lindquist (2026)).  

\subsection{Simulation Results}
\label{sec:simulationresults}
Simulation results with distinctive patterns are presented. Others providing similar conclusion are in Sections 1 of the Supplementary Materials Part II (Zhang, Asgharian and Lindquist (2026)) for $(n,p)=(100,1000)$ and Part III for $(n,p)=(160,480)$, respectively. 

\begin{figure}[b]
\noindent\makebox[\textwidth]{%
		\includegraphics[width=1.15\textwidth]{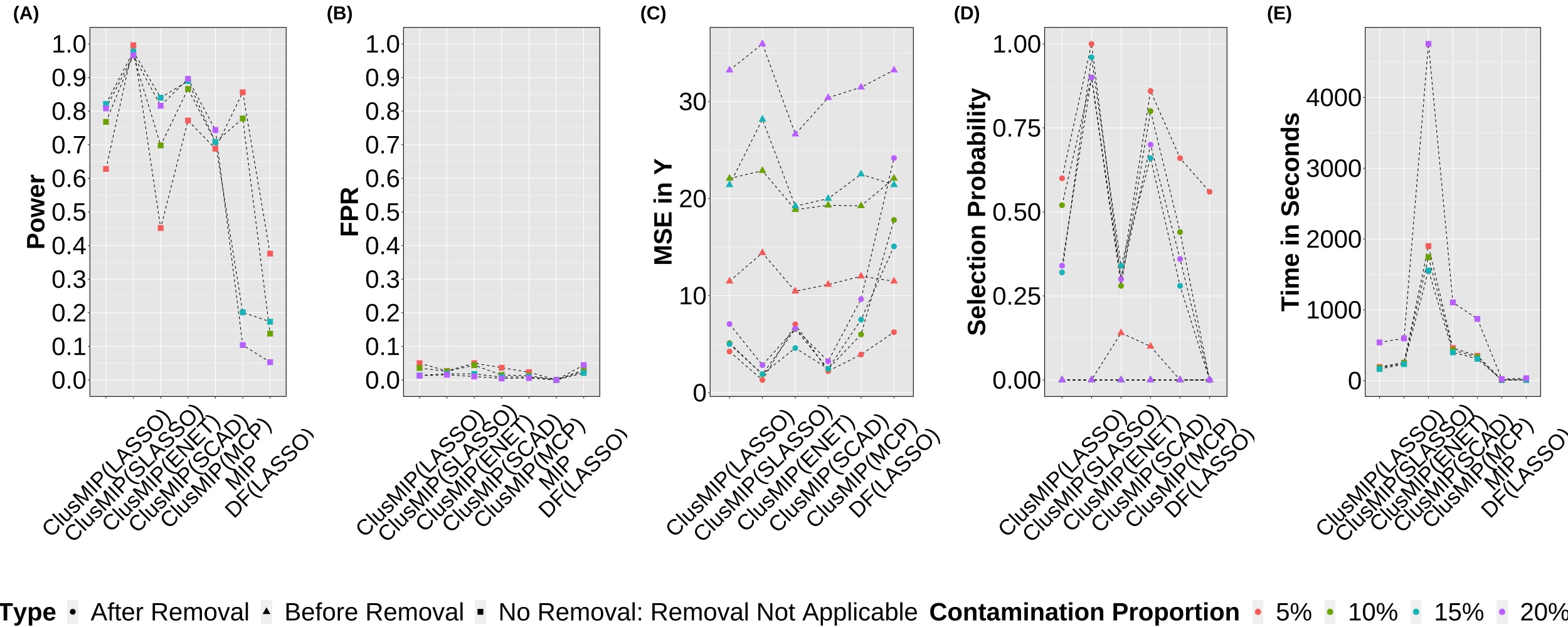}}
	\caption{Perturbation Scheme I with $(n,p)=(100,1000)$,$\kappa_{\text{Linear}, \textbf{Y}}=$5 and $\textbf{X}=\textbf{X}_{\text{Sc(0.8)}}$: (A) and (B) power and FPR of detection; (C) average number of detected outliers impacting the selection of stable signals; (D) algorithm running time in seconds; (E) and (F) MSE in prediction and probability that a selected submodel contains the true model before and after removing detected influential observations.}
	\label{fig:I_Y5_X0_SC8_n100_p1000}
\end{figure}

\paragraph*{\textbf{Perturbation Scheme I (Figure~\ref{fig:I_Y5_X0_SC8_n100_p1000})}} 

\begin{enumerate}
\item \textbf{Power:} $\text{Power}_{\text{ClusMIP}(\text{SLASSO})}$ and $\text{Power}_{\text{MIP}}$ are generally higher (above 0.90) than the other procedures for the same contamination proportion $\zeta$ at moderate and high contamination scales $\kappa_{\text{linear}, \textbf{Y}}$. This implies that SLASSO is sensitive to influential observations on variable selection, serving as a feasible candidate for use together with an outlier detection procedure. At low values of $\kappa_{\text{linear}, \textbf{Y}}$, concerning under-performance was observed under certain scenarios for the MIP. For example, $\text{Power}_{\text{MIP}}$ drops from 0.20 to 0.10 as $\zeta$ increases from $15\%$ to $20\%$; see Panel (A) of Figure~\ref{fig:I_Y5_X0_SC8_n100_p1000}. In comparison, $\text{Power}_{\text{ClusMIP}(\text{SLASSO})}$ at $\zeta=15\%$ and $20\%$ consistently exceeds 0.97. Moreover, within the ClusMIP procedure, the detection powers of ClusMIP applied to LASSO, ENET, SCAD and MCP show higher variation and are generally lower than $\text{Power}_{\text{ClusMIP}(\text{SLASSO})}$, ranging between 0.45 and 0.89 under the particular simulation setting shown in Figure~\ref{fig:I_Y5_X0_SC8_n100_p1000}. In contrast, $\text{Power}_{\text{DF}(\text{LASSO})}$ (< 0.38 in Figure~\ref{fig:I_Y5_X0_SC8_n100_p1000}) is consistently among the lowest due to its leave-one-out design, indicating its limited ability to detect multiple influential points. In summary, ClusMIP(SLASSO) consistently exhibits better detection result than competing procedures, while under-performance of MIP is observed at low contamination intensity. Further, DF(LASSO) appears to be always outperformed in terms of detection.    

\item \textbf{FPR:} All procedures consistently exhibit small FPR  (less than the nominal level $\alpha=0.05$). This phenomenon justifies  Theorem~\ref{thm:clusmip_fpr} of the ClusMIP procedure and Theorem 4 in \cite{zhao2019} in controlling the FPR. Specifically, the choices of high-dimensional clustering discussed in Section~\ref{sec:clusmip} are ef{}fective in separating the clean and the contaminated portions, limiting the possibility for false detection in subsequent steps. 

\item \textbf{Stable Signal Selection}: At each contamination rate $\zeta$, the ClusMIP of all model selectors detects significantly more outliers that impact the selection of true predictors than DF(LASSO)(Panel (C) of Figure~\ref{fig:I_Y5_X0_SC8_n100_p1000}). In other words, the ClusMIP detects a considerably higher number of genuine outliers, while few are identified by DF(LASSO). Specifically, $\text{Stable}_{\text{DF(LASSO)}}=1.95, 1.4, 2.75$ and $1.1$ at $\zeta=5\%, 10\%, 15\%$ and $20\%$ (i.e. at $n_\text{infl}=5,10,15$ and 20). In comparison,  $\text{Stable}_{\text{ClusMIP(SLASSO)}}$ is $6.1, 9.7, 14.9$ and $18.95$. This result, corroborated by the pattern of detection power (Panel (A)), reinforces that ClusMIP outperforms DF(LASSO) in accurately identifying true outliers. 

\item \textbf{Computation Time:} In general, we have $\text{Time}_{\text{MIP}} < \text{Time}_{\text{DF}(\text{LASSO})} < \{\text{Time}_{\text{ClusMIP}(\text{LASSO})}, \\ \ll \text{Time}_{\text{ClusMIP}(\text{SLASSO})}, \text{Time}_{\text{ClusMIP}(\text{SCAD})},  \text{Time}_{\text{ClusMIP}(\text{MCP})} \} \ll \text{Time}_{\text{ClusMIP}(\text{ENET})}$. Noticeably, $\text{Time}_{\text{MIP}}$ and $\text{Time}_{\text{DF}(\text{LASSO})}$ are negligible comparing with that of the ClusMIP approach. This is due to the fact that MIP requires no model fitting. Moreover, DF(LASSO) requires models to be fit $(n+1)$ times. In contrast, the ClusMIP framework requires  models to be fit $|\widehat{\text{S}}_{\text{infl}}|(2+|n-\widehat{\text{S}}_{\text{infl}}|)$ times, where $\widehat{\text{S}}_{\text{infl}}$ is the number of estimated influential points from the high-dimensional clustering scheme. This substantially surpasses that for the DF(LASSO). On the other hand, within the ClusMIP framework, the speed varies according to the types of penalty functions and particular $\tt{R}$ implementation algorithms. Specifically, the ENET includes a two-dimensional exhaustive search of the tuning parameter pair, leading to the largest computation time. In comparison, the SLASSO simultaneously estimates the noise and coef{}ficient, and its $\tt{R}$ package $\tt{scalreg}$ provides faster convergence than other penalized model selectors under consideration. 

\item \textbf{MSE of Prediction:} Lower prediction MSE after removing detected influential observations is consistently observed across all procedures, contamination proportions, and scales. For example, the MSE of ClusMIP(SLASSO) and DF(LASSO) declines from 35.94 to 2.84 and from 33.23 to 24.18, respectively, at $\zeta=20\%$ as seen in Panel (E) of Figure~\ref{fig:I_Y5_X0_SC8_n100_p1000}. While the extent of reduction in MSE varies according to the detection procedure, model selector, and contamination scales, MSE is always reduced after eliminating detected outliers, indicating the benefit of their detection. Moreover, low MSE after outlier removal is linked to improved detection performance. Noticeably, the MSE of SLASSO after the removal of outliers is always among the lowest, due to its strong detection performance. In comparison, the MSE after outlier removal for the remaining ClusMIP procedures showshigher fluctuation across simulation settings but generally corresponds to their detection power. For example, in Panels (A) and (E) of Figure~\ref{fig:I_Y5_X0_SC8_n100_p1000}, the MSE of ClusMIP(SCAD) after outlier removal (between 2.22 and 3.24) is the second lowest which agrees with the second-best position of its detection power performance (between 0.77 and 0.90). In contrast, the MSE of the DF(LASSO) after outlier removal is higher than that of the ClusMIP procedures due to its worse outlier detection. 

\item \textbf{Variable Selection}: Improvement in the empirical probability that the true model is included in the selected submodel after removing the detected influential points is generally noted for all the procedures, indicating that these observations are indeed influential on variable selection. Similar to the reduction in MSE discussed above, the performance of this assessment metric hinges on the detection outcome. Due to reduction in the heterogeneity in the data, better detection performance contributes to higher probability of choosing a submodel containing the true model after outlier removal. This is analogous to results seen for detection power. This is numerically corroborated by comparing Panels (A) and (F) of Figure~\ref{fig:I_Y5_X0_SC8_n100_p1000}: ClusMIP(SLASSO) attains the highest detection power (0.97 -- 0.99), followed by ClusMIP(SCAD) (0.77 -- 0.90). On the other hand, while the selection probability is approximately 0 for both before outlier removal (i.e., based on the contaminated dataset), the two procedures achieve the highest selection probabilities after outlier removal, ranging between 0.90 -- 1.00 and 0.66 -- 0.86, respectively. In contrast, the selection probability after outlier removal for DF(LASSO) is 0 at $\zeta=10\%$, $15\%$ and $20\%$, which can be explained by its corresponding  poor performance in detection. 
\end{enumerate}

\begin{figure}[b]
\noindent\makebox[\textwidth]{%
		\includegraphics[width=1.1\textwidth]{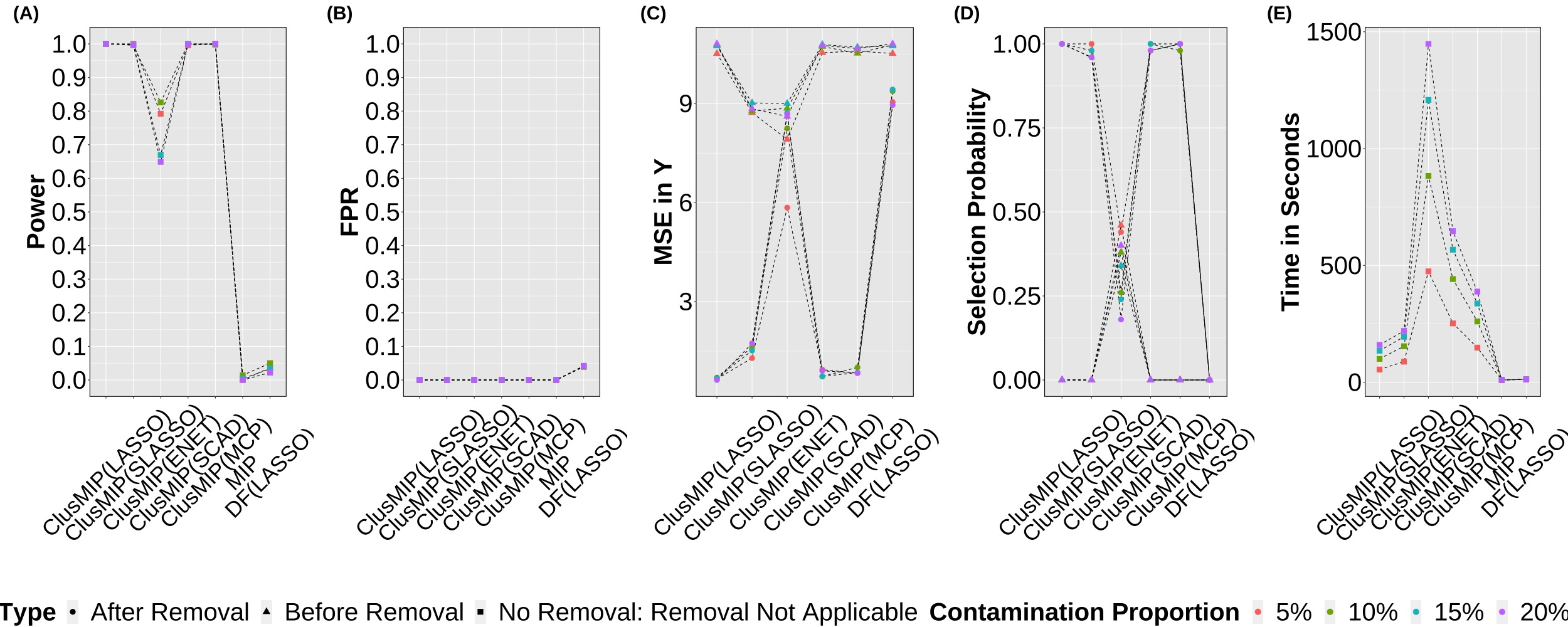}}
	\caption{Perturbation Scheme II with (n,p)=(100,1000), $\kappa_{\text{Linear}, \textbf{X}}$=10 and $\textbf{X} = \textbf{X}_{\text{Sc(0.8)}}$: (A) and (B) power and FPR of detection; (C) average number of detected outliers that impact the selection of stable signals; (D) algorithm running time in seconds; (E) and (F) MSE in prediction and probability that a selected submodel contains the true model before and after removing detected influential observations.}
	\label{fig:II_Y0_X10_SC8_n100_p1000}
\end{figure}

\paragraph*{\textbf{Perturbation Scheme II (Figure~\ref{fig:II_Y0_X10_SC8_n100_p1000})}}

\begin{enumerate}
    \item \textbf{Power:} The detection power of the two existing methods, MIP and DF(LASSO) are no more than 0.10, considerably falling behind those of the newly proposed ClusMIP procedure across all simulation settings. While the poor performance of DF(LASSO) is dues to  its leave-one-out design, for MIP, we postulate that contaminating the predictors while keeping the responses intact is unlikely to impose sufficient impact on the marginal correlation-based HIM measure, leading to lower detection power. 

    When $\kappa_{\text{Linear}, \textbf{X}}$ is moderate and for all types of design matrices,  $\text{Power}_{\text{ClusMIP}(\text{LASSO})}$, $\text{Power}_{\text{ClusMIP}(\text{SLASSO})}$, $\text{Power}_{\text{ClusMIP}(\text{SCAD})}$ and $\text{Power}_{\text{ClusMIP}(\text{MCP})}$ perform best, and all approximately exceed 0.90. In contrast, $\text{Power}_{\text{ClusMIP}(\text{ENET})}$ is the worst performer in the ClusMIP family, showing larger variation over $\zeta$. As seen in Panel (A) of Figure~\ref{fig:II_Y0_X10_SC8_n100_p1000}, $\text{Power}_{\text{ClusMIP}(\text{ENET})}$ ranges between 0.65 and 0.85 in that specific simulation setting. One contributing factor is the need to select two tuning parameters for ENET may complicate and impede the formation of the GDF measure \eqref{gdf} to accurately exhibit the ef{}fects of influential observations. Indeed,  poor performance of ClusMIP(ENET) in detection seems to be observed across all the simulation settings. 

    \item \textbf{MSE of Prediction and Variable Selection:} Similar to the conclusion from Perturbation Scheme I, the MSE of prediction and selection probability after removing outliers depends on the detection performance. Specifically, better detection results lead to lower MSE and higher selection probability after outlier removal. For example, as seen in Panels (C) and (D) of Figure~\ref{fig:II_Y0_X10_SC8_n100_p1000}, severe under-performance of DF(LASSO) in detection causes the highest MSE (between 8.96 and 9.42) and lowest selection probability (always 0). In contrast, ClusMIP(SCAD) and ClusMIP(SLASSO) attain the lowest MSE (between 0.73 and 0.94, and 1.29 and 1.73, respectively) and the highest selection probability (between 0.98 and 1.00, and 0.96 and 1.00, respectively) after removal of detected influential points. 

    \item \textbf{FPR, Stable Signal Selection, and Computation Time:} they are similar to those under Perturbation Scheme I.  
\end{enumerate}

\paragraph*{\textbf{Perturbation Scheme III}} 
The patterns of detection power, FPR, computation time, MSE of prediction and selection probability are analogous to the ones of Perturbation Scheme I. 

\section{Analysis of Neurologic Signature of Physical Pain} 
\label{sec:realdata}

We apply existing procedures, as well as the newly proposed ClusMIP procedures to the fMRI data described in Section~\ref{sec:real_background}. We stress that even though the study \citep{lindquist2017} focuses on pain prediction, the pitfalls of ignoring outliers on downstream analysis have been thoroughly discussed both from scientific angles and via a synthetic example in Section~\ref{sec:introduction}. In particular, we have shown that the presence of multiple outliers significantly distorts variable selection and mitigates predictive power. Therefore, we focus on detecting influential points in Section~\ref{sec:real_detection}. The benefits of excluding the identified outliers on variable selection, coef{}ficient estimation, and prediction are then systematically examined in Sections~\ref{sec:real_benefits_variable_selection} to \ref{sec:real_benefits_prediction}. Based on our analyses, we conclude by providing concrete advice on the practical use of the newly proposed procedure for application to the thermal pain dataset \citep{lindquist2017} in Section~\ref{sec:real_advice}. 

Two metrics are considered to assess detection outcome: $\widehat{\text{I}}^{\text{infl}}_{\text{Detection}(\text{Selector})}$ and $\vert \widehat{\text{I}}^{\text{infl}}_{\text{Detection}(\text{Selector})}\rvert$. They refer to the estimated index set of influential points and its cardinality, respectively. To obtain them, the existing methods, MIP and DF(LASSO), and the ClusMIP procedure applied to LASSO, SLASSO, ENET, SCAD and MCP are performed. When computing these two assessment measures, the tuning parameter is chosen as follows: for LASSO, ten-fold CV is used; for SLASSO, the tuning parameter is fixed as $\sqrt{(2\log(p)/n)}$ \citep{sun2012}; for ENET, two-dimensional ten-fold CV is used to select the tuning parameter pair; for SCAD and MCP, the thresholding parameters are set to 3.7 and 3.0, respectively \citep{ncvreg}, while the other parameter controlling the scale of shrinkage is determined via ten-fold CV. Here, we recognize that the selection of tuning parameter(s) via $k$-fold CV is stochastic and this randomness may impact the calculation of detected influential points. This necessitates mitigation of the embedded randomness, leading to the detection of more reliable and reproducible outliers. One remedial procedure is repeated $k$-fold CV, where $k$-fold CV is repeated a specified number of times. According to our simulation study provided in Section 13 of the Supplementary Materials Part I (Zhang, Asgharian and Lindquist (2026)), the ClusMIP procedure used with the model selectors under consideration exhibits robustness to data-dependent influential point detection. To be more specific, detection outcomes (measured via the power and FPR of detection, proportion of each individual datum labeled as influential and average number of outliers impacting the selection of true predictors) based on the classic (single) ten-fold CV are almost identical to those obtained when using repeated ten-fold CV. As such, the outliers for all model selectors are obtained using a single ten-fold CV (when applicable) and treated as deterministic for subsequent analysis. Values for these two metrics when using different detection procedures are reported in Table~\ref{tab:pain_prediction_detection}. 

To quantify the benefits of detecting these influential points, each regularized model selectors are fit to both the full and reduced datasets, where the latter dataset is obtain after removing outliers. Considering the random nature of ten-fold CV in tuning parameter selection (except for SLASSO), in contrast to influential point detection, the downstream analyses on variable selection, coef{}ficient estimation, and prediction based on using both the full and reduced datasets are repeated 1000 times. The empirical averages of these associated benchmarks (defined in  Sections~\ref{sec:real_benefits_variable_selection} to \ref{sec:real_benefits_prediction}) are  reported for comparison purposes. Since LASSO is used for both DF(LASSO) and ClusMIP(LASSO), to avoid ambiguity, results for both procedures are shown separately in Figures~\ref{fig:real_plot_aoas_revision_effects_detection_clusmip}
 and \ref{fig:real_plot_aoas_revision_effects_detection_dflasso}.

\subsection{Detection of Influential Observations}
\label{sec:real_detection}

From Table~\ref{tab:pain_prediction_detection}, we observe that dif{}ferent combinations of detection procedures and model selectors cause drastically distinctive detection outcomes. Specifically, while no points are labeled as influential by ClusMIP(SLASSO) and MIP, more than 16 observations are deemed influential by ClusMIP(LASSO), ClusMIP(ENET), ClusMIP(SCAD) and ClusMIP(MCP). This substantiates our previous explanation in Section~\ref{sec:principles} that the detection of influential observations on variable selection is intricately linked to the properties of variable selection procedures. To be more specific, dif{}ferent penalty functions of the model selectors reveal dif{}ferent sensitivities with respect to variable selection in the presence of influential observations. For this particular dataset, LASSO seems to be the most sensitive model selector, followed by SCAD, MCP and ENET. 

\begin{table}[t]
\centering
 \addtolength{\leftskip}{-2cm}
    \addtolength{\rightskip}{-2cm}
    \fontsize{8}{11}\selectfont
    \caption{\centering Assessment Metrics for the Pain Prediction Dataset \citep{lindquist2017}}
\begin{tabular}{|C{2.4cm}|C{8.1cm}|C{2.5cm}|}
\hline 
Detection(Selectors) & \makecell{$\widehat{\text{I}}^{\text{infl}}_{\text{Detection(Selector)}}$} & \makecell{$\big\vert \widehat{\text{I}}^{\text{infl}}_{\text{Detection (\text{Selector}})} \big \rvert$} \\ \hline 
\makecell{ClusMIP(LASSO)}  & 3, 7, 13, 14, 15, 19, 22, 25, 26, 43, 44, 49, 50, 52, 55, 56, 57, 58, 62, 67, 75, 79, 80, 81, 103, 115, 122, 123, 124, 127, 145, 147, 157, 158, 159, 164, 165, 169, 170, 172, 175, 176,
194 & 42  \\ \hline
\makecell{ClusMIP(SLASSO)} & N.A.& 0  \\ \hline
\makecell{ClusMIP(ENET)} & 31, 49, 50, 52, 55, 58, 76, 79, 115, 122, 124, 158, 169, 170, 172, 194 & 16 \\ \hline
\makecell{ClusMIP(SCAD)} & 31, 43, 49, 52, 55, 57, 62, 76, 80, 86, 103, 121, 133, 134, 145, 146, 157, 159, 164, 165, 173, 177, 181, 182, 193 & 25 \\ \hline
\makecell{ClusMIP(MCP)} & 13, 19, 26, 49, 52, 55, 57, 58, 73, 76, 86, 103, 121, 124, 134, 145, 147, 151, 163, 164, 165, 182 & 22  \\ \hline
MIP & N.A. & 0 \\ \hline
\makecell{DF(LASSO)} & 2, 57, 84, 119, 153, 159, 168, 174, 198 & 9  \\ \hline
\end{tabular}
\label{tab:pain_prediction_detection}
\end{table}

For ClusMIP(LASSO) and ClusMIP(ENET), we observe that $\widehat{\text{I}}^{\text{infl}}_{\text{ClusMIP(ENET)}}$ is almost a proper subset of $\widehat{\text{I}}^{\text{infl}}_{\text{ClusMIP(LASSO)}}$, except for observations 31 and 76. Indeed, $\lvert \widehat{\text{I}}^{\text{infl}}_{\text{ClusMIP(LASSO)}} \rvert$ is the largest among all procedures considered. One possible explanation is that for this dataset, predictors which are multivariate brain activity patterns represented by voxels are expected to be highly correlated. Therefore, certain significant predictors may fail to be selected by LASSO. On the other hand, the use of both $\ell_1$ and $\ell_2$ penalties in ENET facilitates the improved performance in variable selection. Thus, the submodel selected by LASSO may be more parsimonious than the one selected by ENET. This necessarily implies that in the finite sample, the submodel selected by ENET may demonstrate better goodness-of-fit than the one chosen by LASSO. As such, in the presence of influential observations, LASSO may have a higher tendency to be impacted by perturbation in variable selection than  ENET. Finally, the set $\widehat{\text{I}}^{\text{infl}}_{\text{DF}(\text{LASSO})}$ exhibits the lowest cardinality and hardly overlaps with those identified by other procedures, except for observation 57. In fact, this low cardinality is consistent with the simulation studies, where the detection power of DF(LASSO) is always among the lowest (Panel (A) of Figures~\ref{fig:I_Y5_X0_SC8_n100_p1000} and \ref{fig:II_Y0_X10_SC8_n100_p1000}). 

\begin{figure}[b]
\noindent\makebox[\textwidth]{%
		\includegraphics[width=1.0\textwidth]{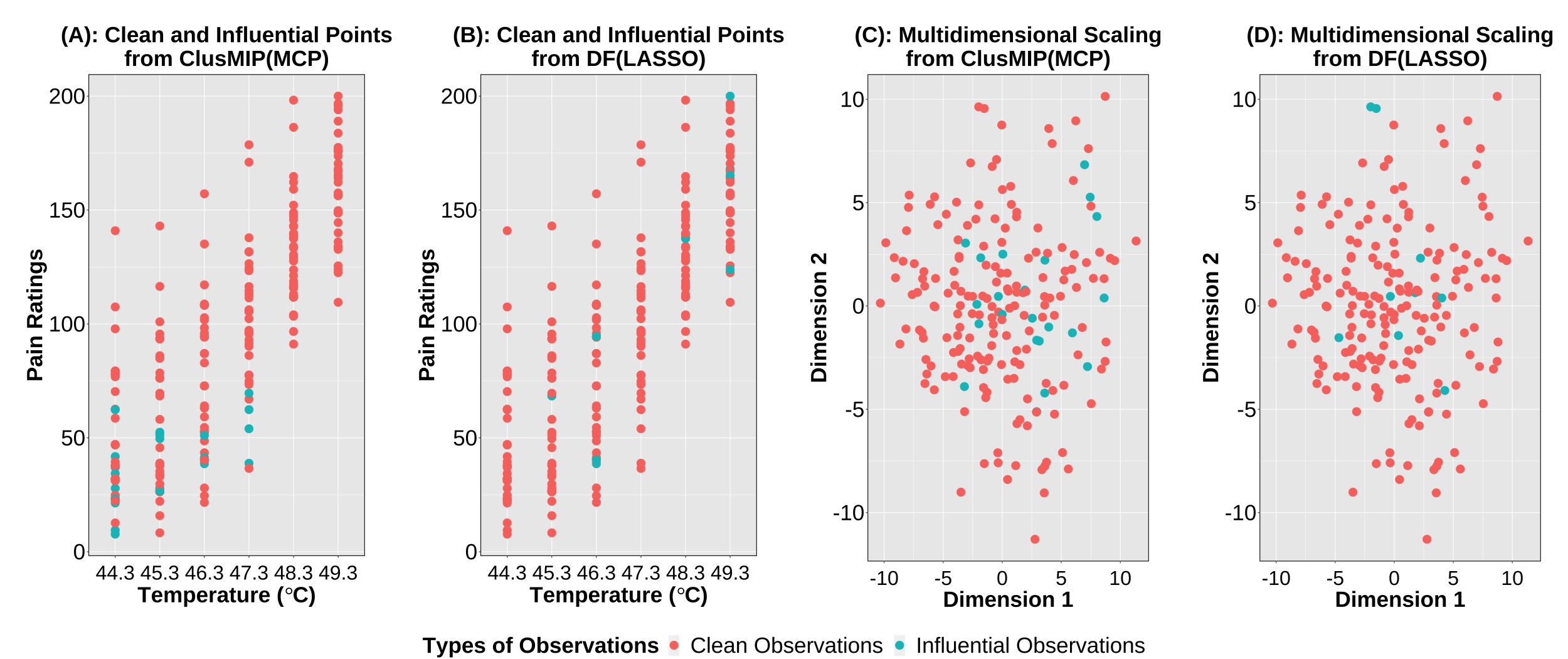}}
	\caption{(A) and (B): respective distributions of $\widehat{\text{I}}^{\text{infl}}_{\text{ClusMIP}(\text{MCP})}$ and $\widehat{\text{I}}^{\text{infl}}_{\text{DF(LASSO)}}$. (C) and (D): the first two dimensions from multi-dimensional scaling respectively from the ClusMIP(MCP) and DF(LASSO).}
	\label{fig:real_data_analysis_detection}
\end{figure} 

Furthermore, we observe that $  \widehat{\text{I}}^{\text{infl}}_{\text{Common}} \coloneqq \ 
     \widehat{\text{I}}^{\text{infl}}_{\text{ClusMIP}(\text{LASSO})} \ \cap \ 
     \widehat{\text{I}}^{\text{infl}}_{\text{ClusMIP}(\text{ENET})}  \ \cap \
     \widehat{\text{I}}^{\text{infl}}_{\text{ClusMIP}(\text{SCAD})}  \ \cap \ 
     \widehat{\text{I}}^{\text{infl}}_{\text{ClusMIP}(\text{MCP})}   = \{49, 52, 55\}$.     
Hence, observations 49, 52 and 55 are flagged for further investigation. The first two observations correspond to two different temperatures measured on the same participant, indicating this may be a participant that needs to be further investigated. In addition, the first and third observations correspond to the lowest temperature, possibly indicating participants that have a differing pain threshold than other participants. Indeed, a plot of the pain ratings as function of temperature with influential points identified by the ClusMIP(MCP) (Figure~\ref{fig:real_data_analysis_detection}(A)) indicates that the majority of outliers correspond to low pain ratings at each temperature, a pattern which holds for ClusMIP across all model selectors. For illustration, only the plot for ClusMIP(MCP) is shown (Figure~\ref{fig:real_data_analysis_detection}(A)) while the other plots in the ClusMIP family are presented in Section 12 of the Supplementary Materials Part I (Zhang, Asgharian and Lindquist (2026)). In fact, the low pain score primarily contributes to the ``outlyingness'' of the identified influential points, as no anomalies in the feature space are detected as seen in the multi-dimensional scaling plot (Figure~\ref{fig:real_data_analysis_detection}(C)). This may be explained by the fact that this temperature is generally considered warm, but not always painful, thus dif{}ferentiating it from other temperature levels. Together these two panels imply that the low pain ratings may need to be described using a dif{}ferent model than that of high pains in fMRI-based thermal pain modeling. This assertion is further strengthened by research on pain processing mechanisms suggesting a discrepancy in brain activity between low and intense pain conditions \citep{fairhurst2012}. In contrast, the outliers of DF(LASSO) do not exhibit an easily recognizable pattern (Figure~\ref{fig:real_data_analysis_detection}(B)), consisting of both low and high pain scores. This may be due to its limited ability to detect multiple outliers, causing challenges for meaningful interpretation and reliable inference. 

\subsection{Ef{}fects of Influential Observations on Variable Selection}
\label{sec:real_benefits_variable_selection}

\begin{figure}[t]
\noindent\makebox[\textwidth]{%
		\includegraphics[width=1.0\textwidth]{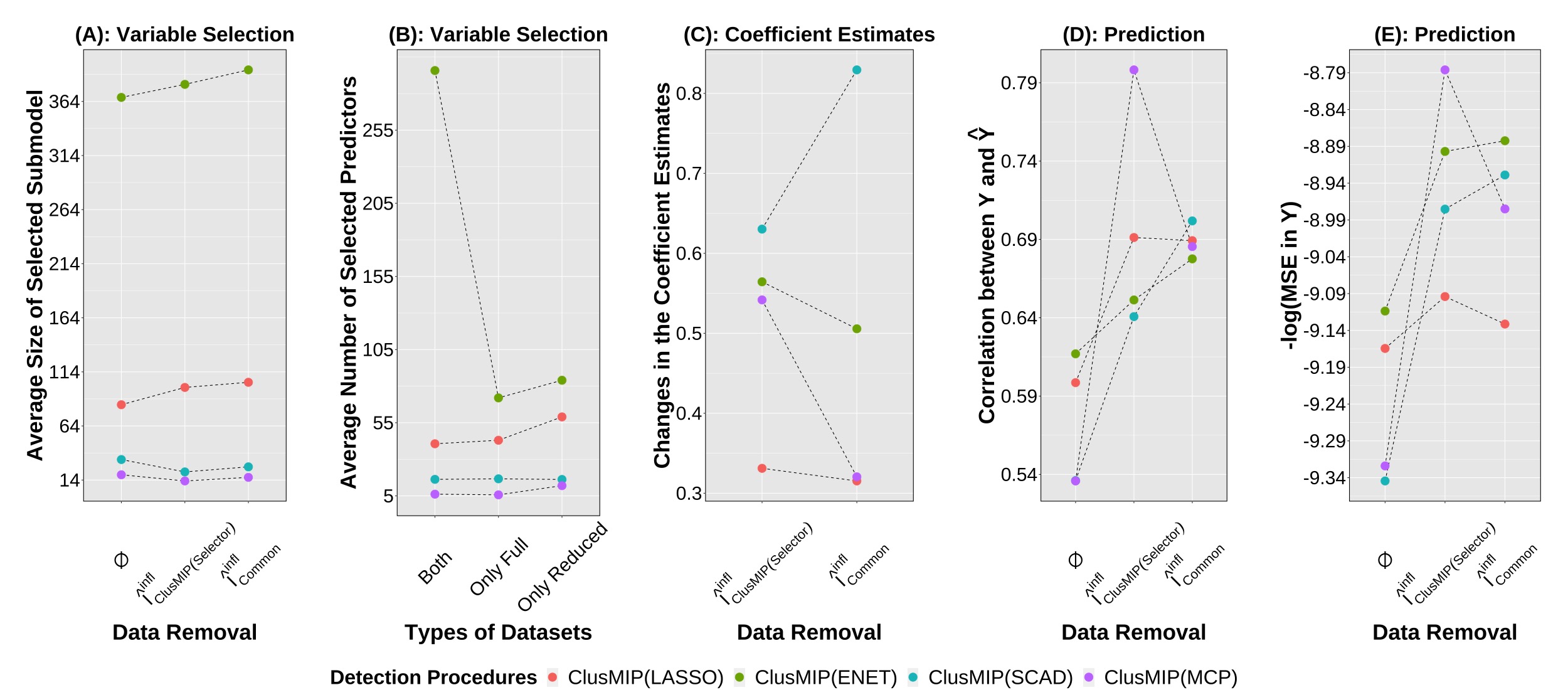}}
	\caption{ClusMIP. (A): average sizes of submodels selected from the full and reduced datasets upon removing $\widehat{\text{I}}^{\text{infl}}_{\text{ClusMIP(Selector)}}$ and $\widehat{\text{I}}^{\text{infl}}_{\text{Common}}$. (B): average number of predictors included in submodels selected from both full and reduced datasets, unique to those from the full dataset, and unique to those from the reduced dataset. (C): change in the normed coefficient estimates upon outlier removal. (D): sample correlation between observed and predicted pain scores. (E): $-\log{(\text{MSE of Prediction})}$.}
\label{fig:real_plot_aoas_revision_effects_detection_clusmip}
\end{figure}

\begin{figure}[t]
\noindent\makebox[\textwidth]{%
		\includegraphics[width=1.0\textwidth]{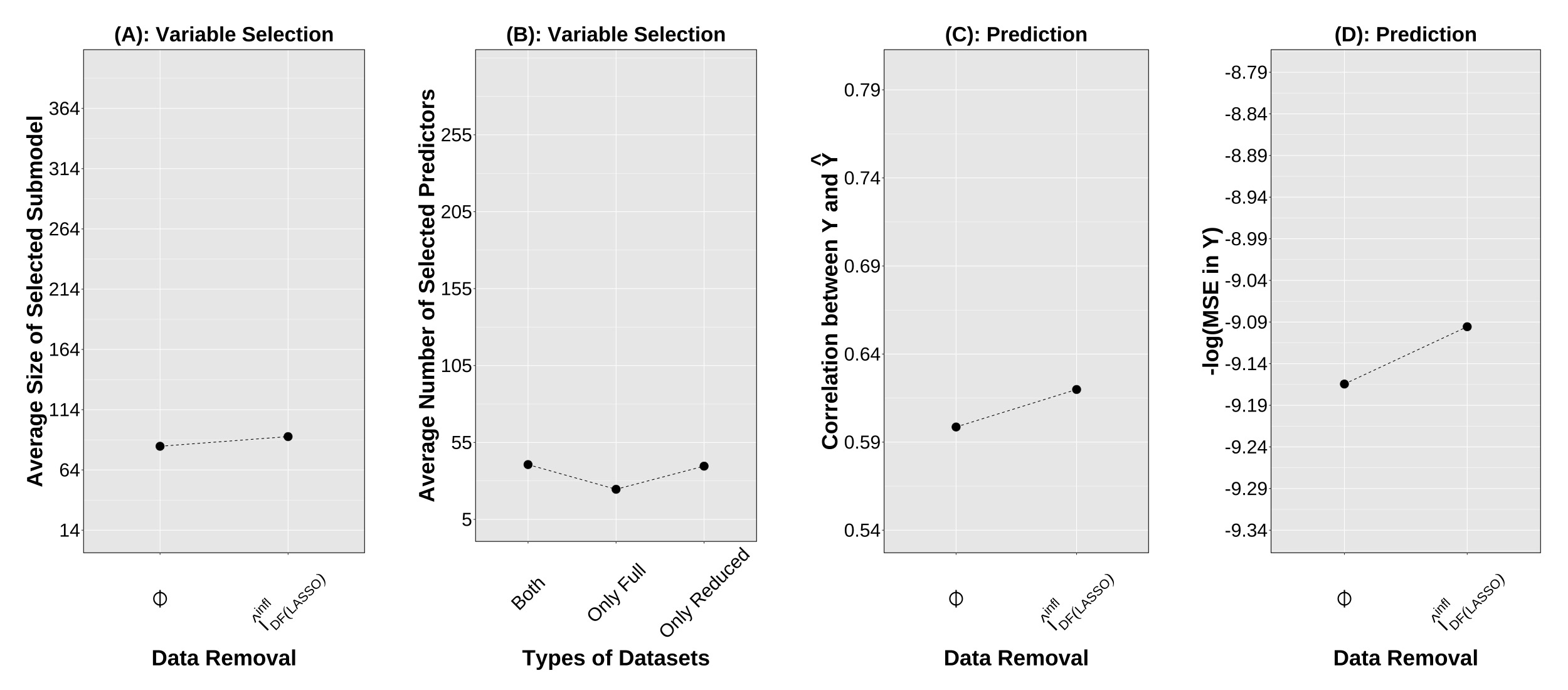}}
	\caption{DF(LASSO). (A): average size of submodels chosen from the full and reduced datasets (removing $\widehat{\text{I}}^{\text{infl}}_{\text{DF(LASSO)}}$). (B): average number of predictors that are in the submodels selected from both full and reduced datasets, unique to those from the full dataset, and unique to those from the reduced dataset. (C): sample correlation between observed and predicted pain scores. (D): $-\log{(\text{MSE of Prediction})}$.}
\label{fig:real_plot_aoas_revision_effects_detection_dflasso}
\end{figure}

The detected influential points demonstrate non-negligible impact on variable selection.
For ClusMIP, this is assessed by computing the average size of submodels selected from the following three datasets: the full dataset and the two reduced datasets upon removing $\widehat{\text{I}}^{\text{infl}}_{\text{ClusMIP(Selector)}}$ and $\widehat{\text{I}}^{\text{infl}}_{\text{Common}}$; see Figure~\ref{fig:real_plot_aoas_revision_effects_detection_clusmip}(A). For DF(LASSO), equivalent results, shown in Figure~\ref{fig:real_plot_aoas_revision_effects_detection_dflasso}(A), are obtained from the full dataset and only one reduced dataset obtained by removing $\widehat{\text{I}}^{\text{infl}}_{\text{DF(LASSO)}}$ since $\widehat{\text{I}}^{\text{infl}}_{\text{Common}}$ does not overlap with $\widehat{\text{I}}^{\text{infl}}_{\text{DF(LASSO)}}$. 

In Panel (A) of Figures~\ref{fig:real_plot_aoas_revision_effects_detection_clusmip} and \ref{fig:real_plot_aoas_revision_effects_detection_dflasso}, variation in the average model sizes before and after outlier removal is clearly seen. Specifically, such quantities for LASSO, ENET, SCAD and MCP before outlier removal via ClusMIP are 83.63, 367.72, 32.95 and 18.87, which change to 99.61, 379.78, 21.51 and 13.16 after the removal. Moreover, for LASSO, the model size increases from 83.63 to 91.65 upon removing those identified via DF(LASSO). Furthermore, while the model sizes increase after outlier removal for the two convex penalties LASSO and ENET, those selected by the two non-convex penalties SCAD and MCP are more sparse after removal of detected outliers. Also, LASSO and ENET select considerably more predictors than SCAD and MCP both before and after eliminating identified influential observations. 

To supplement Panels (A) of Figures~\ref{fig:real_plot_aoas_revision_effects_detection_clusmip} and \ref{fig:real_plot_aoas_revision_effects_detection_dflasso}, we also calculate the average number of common predictors that are contained in submodels selected from both full and reduced datasets, and those that are unique to the submodels chosen from the full and reduced datasets, shown in Panels (B) of Figures~\ref{fig:real_plot_aoas_revision_effects_detection_clusmip}  and \ref{fig:real_plot_aoas_revision_effects_detection_dflasso}, respectively. From these plots, a significant number of predictors unique only to the reduced datasets are seen for all model selectors. Specifically, 58.99, 84.04, 16.22 and 12.00 unique predictors are chosen by LASSO, ENET, SCAD and MCP based on the reduced datasets obtained by removing $\widehat{\text{I}}^{\text{infl}}_{\text{ClusMIP(Selector)}}$, consisting of $59.22\%$, $22.12\%$, $75.40\%$ and $91.19\%$ of their respective  submodels. Moreover, LASSO selects 39.60 unique predictors upon removing $\widehat{\text{I}}^{\text{infl}}_{\text{DF(LASSO)}}$, $43.21\%$ of its submodel size. These phenomena further 
indicates the impact of influential observations on variable selection. 

In Figure~\ref{fig:real_brain_plot}, we illustrate the ef{}fects of influential points on variable selection by highlighting the brain regions selected via MCP with color intensities represented by the corresponding coef{}ficient estimates obtained from the full and reduced datasets. Panels A and B show the discrepancy in selected regions, while Panels C, D and E pinpoint common and uniquely selected regions from the two datasets. Six regions are common to both models, including primary visual, frontoparietal, dorsal attention, and default mode areas, as well as the amygdala. Regions unique to the full model include the cerebellum, while regions unique to the reduced model include ventral attention and additional frontoparietal areas. The frontoparietal network, dorsal attention network, visual networks and the amygdala are all associated with pain processing, justifying their inclusion in both models. On the other hand, the ventral attention network (VAN) is responsible for processing unexpected external stimuli. In this study, stronger responses resulting from the high pain scores may lead to more robust bottom-up activation of the VAN. This justifies the inclusion of the VAN in the model chosen from the reduced dataset after removal of the ``outlying'' low pain scores. 

\subsection{Ef{}fects of Influential Observations on Coef{}ficient Estimation}
\label{sec:real_benefits_coefficient_estimation}

In Figure~\ref{fig:real_plot_aoas_revision_effects_detection_clusmip}(C), we examine the impact of influential points on coefficient estimation by computing the percentage change in the $\ell_2$-normed  coefficient estimates defined by $(||\widehat{\boldsymbol{\beta}}_{\text{full}}||_2-||\widehat{\boldsymbol{\beta}}_{\text{redu}}||_2)/||\widehat{\boldsymbol{\beta}}_{\text{full}}||_2$, where $\widehat{\boldsymbol{\beta}}_{\text{full}}$ and $\widehat{\boldsymbol{\beta}}_{\text{redu}}$ are the coef{}ficient estimates obtained by fitting the full dataset and reduced dataset via removing $\widehat{\text{I}}^{\text{infl}}_{\text{ClusMIP(Selector)}}$ or $\widehat{\text{I}}^{\text{infl}}_{\text{Common}}$. From Figure~\ref{fig:real_plot_aoas_revision_effects_detection_clusmip}(C), at least $30\%$ change in the $\ell_2$-normed coef{}ficients is seen for all model selectors from these two reduced datasets. The largest changes, $63.03\%$ and $82.94\%$, are observed for SCAD. In comparison, although LASSO shows the smallest coef{}ficient adjustments, they reach significant levels of $33.11\%$ and $31.53\%$. The same benchmark is also calculated for DF(LASSO) upon removing $\widehat{\text{I}}^{\text{infl}}_{\text{DF(LASSO)}}$ only, which attains $34.22\%$. These results highlight the considerable impact of outliers on coefficient estimation.

\subsection{Ef{}fects of Influential Observations on Prediction}
\label{sec:real_benefits_prediction}

Improved prediction power is also observed upon removal of detected influential observations. This is assessed using two criteria: the sample correlation between observed and predicted pain scores and the logarithm of MSE for prediction. They are shown separately in Figure~\ref{fig:real_plot_aoas_revision_effects_detection_clusmip}(D) and (E) for ClusMIP, and Figure~\ref{fig:real_plot_aoas_revision_effects_detection_dflasso}(C) and (D) for DF(LASSO). Clear improvements are seen after removing detected influential observations for all model selectors. For example, the highest sample correlation is obtained by MCP, which increases from 0.53 to 0.80 after removing $\widehat{\text{I}}^{\text{infl}}_{\text{ClusMIP(MCP)}}$. This is accompanied by improved performance in the MSE of prediction, which declines from $\exp{(9.32)}=11203.49$ to $\exp{(8.79)}=6542.97$, the largest such reduction ($41.6\%$) among all model selectors. In contrast, no significant dif{}ference of sample correlation is seen for LASSO upon removing the outliers detected by DF(LASSO) (Figure~\ref{fig:real_plot_aoas_revision_effects_detection_dflasso}(C)). In contrast, the MSE of LASSO drops from $\exp{(9.16)}=9551.72$ to $\exp{(9.10)}=8916.39$ after removing the outliers detected by DF(LASSO), which is $36.27\%$ higher than that obtained after removing the outliers detected by ClusMIP(MCP). Thus, the difference in prediction power before and after outlier elimination illustrates the benefits of carefully considering points identified by all detection procedures. Moreover, the advantage of ClusMIP over DF(LASSO) in terms of predictive performance is clearly observed.

\begin{figure}[t]
\noindent\makebox[\textwidth]{%
		\includegraphics[width=0.9\textwidth]{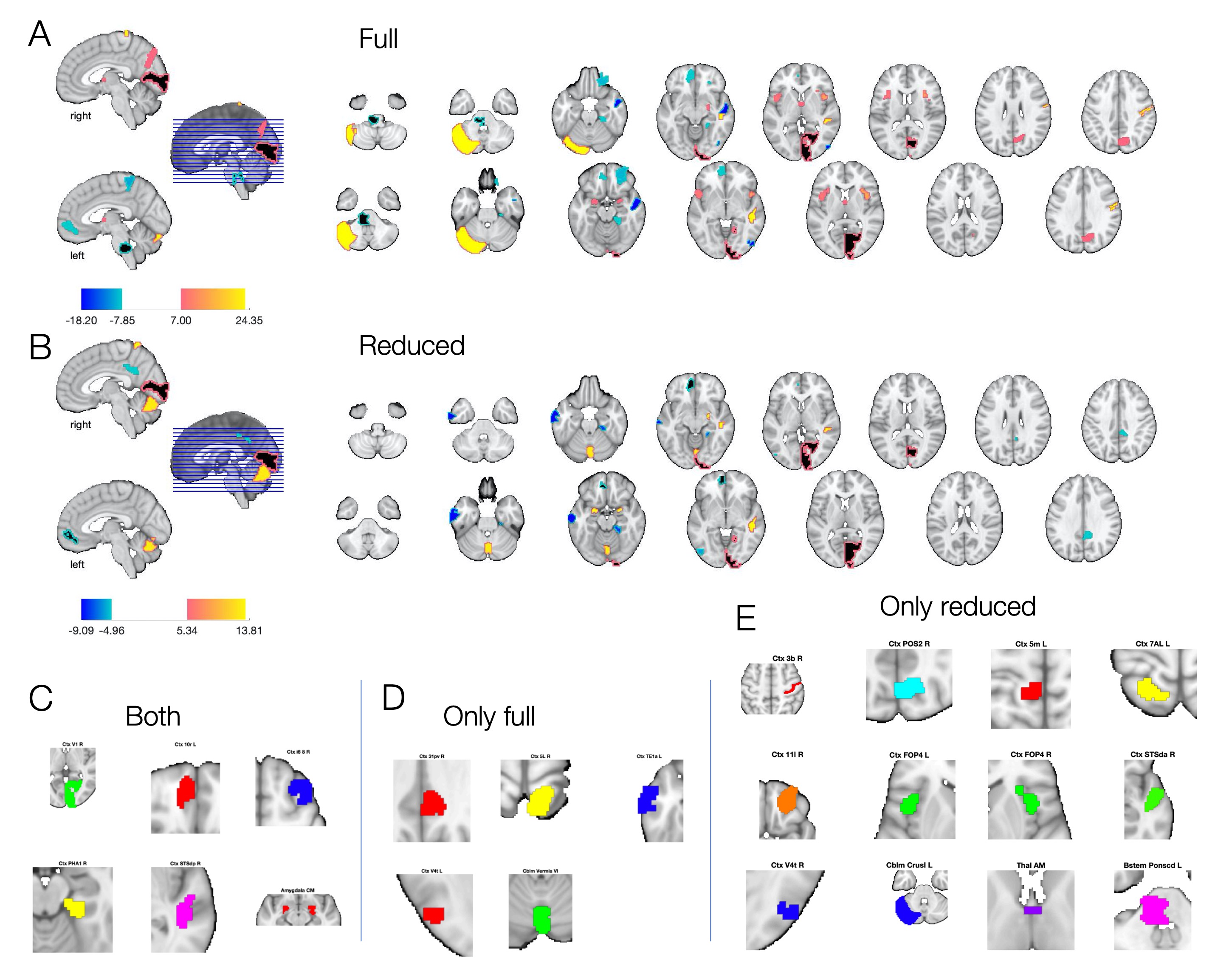}}
	\caption{(A) and (B): brain regions selected by MCP from the full and reduced datasets with color intensities denoting the scales of coef{}ficient estimates; (C): common regions selected from both datasets; (D) and (E): unique regions selected respectively from the full and reduced dataset.}
	\label{fig:real_brain_plot}
\end{figure} 

\subsection{Summary of Analysis}
\label{sec:real_advice}

In developing a neurologic signature of physical pain, the principal objective is to create a single model predictive of pain levels. It is critical to remove influential observations specific to the particular model selector used for deriving the model. In the context of our thermal pain data, the influential observations detected by ClusMIP(MCP) correspond to low pain scores (Figure~\ref{fig:real_data_analysis_detection}(A)) for each temperature. This pattern highlights the distinction between low and intense pain processing mechanisms, consistent with findings in other pain studies. Moreover, after exclusion of these points, MCP selects a statistical model containing unique feature(s) (the VAN network) (Figure~\ref{fig:real_brain_plot}), which regulates the responses corresponding to high pain scores. Furthermore, the reduced model selected by MCP is the most sparse (Figure~\ref{fig:real_plot_aoas_revision_effects_detection_clusmip}(A)), yet delivers the best prediction power (Figure~\ref{fig:real_plot_aoas_revision_effects_detection_clusmip}(D) and (E)) among all competing selectors. Specifically, it demonstrates the highest correlation between observed and predicted pain scores, and the lowest MSE in prediction. In other words, upon removing $\widehat{\text{I}}^{\text{infl}}_{\text{ClusMIP(MCP)}}$, MCP selects the most parsimonious, scientifically justifiable, and interpretable model while exhibiting the highest predictive performance. Therefore, we posit that the best performing detection procedure for the thermal pain dataset is ClusMIP(MCP).

\section{Conclusion and Further Discussion}
\label{sec:conclusion}

In this manuscript, we have described the existence of multiple factors that can negatively influence the development of a neurologic signature of physical pain \citep{lindquist2017}, including inter-subject variability in pain processing and functional topology, and the noisy nature of fMRI data. Together these factors can hinder the formation of a generalizable predictive model, leading to concerns about the validity of subsequent statistical inference. 
We illustrate how these factors can  unduly af{}fect the choice of a stochastically selected submodel, an issue that is largely ignored in the existing influential diagnosis literature. In our simulation study, we observe that failure to account for outliers leads to non-negligible, detrimental impacts on various aspects of downstream statistical analysis, including distorted variable selection outcomes, biased coef{}ficient estimates and worse prediction performance. One existing method DF(LASSO) falls short in its theoretical foundation and is not designed to identify multiple influential observations. The inability to detect a block of outliers also holds for the other existing method MIP under certain common contamination scenarios. As such, the existence of outliers in fMRI data, the pitfalls of ignoring them, and the limitations of existing methods necessitate the development of a theoretically sound detection method capable of identifying multiple influential points. 

To guide the development of our detection method, we first recognize that variable selection is pervasive across an 
extensive scope of scientific disciplines for both explanatory and predictive modeling. Moreover, dif{}ferent high-dimensional model selectors (i.e., penalties) have distinctive theoretical properties designed to handle dif{}ferent data types. In fact, in fMRI research, regularized model selectors with both convex and non-convex penalties have been widely applied. Thus, although the motivation for this work originates from the aforementioned thermal pain study, we target a universally adaptable detection framework applicable well beyond the specific application. Consequently, we introduce the following set of guiding principles: (\emph{i}) the ability to detect multiple influential observations; (\emph{ii}) applicability to dif{}ferent model selectors and data types; and (\emph{iii}) theoretical justification.  

One natural question based on our guiding principles is how to interpret  identified influential points, as different model selectors may well lead to different sets of outliers. 
We believe this ultimately depends on the situation. Therefore, we deliberately put forward a spectrum of detection approaches corresponding to different model selectors. The choice of one or multiple model selector(s) hinges on elements which include, but are not limited to, the purpose of study and domain science expertise. Consequently, it is the practitioner' prerogative to decide on the final influential points, ranging from the intersection to the union of all the possible influential point sets. In our case, the goal is to determine a model predictive of pain scores. Precise instruction on the model selector and detection procedure is of{}fered based on the interpretability of detected data points and reduced model, and its resulting performance. 

Using these guiding principles, the GDF measure was proposed in \eqref{gdf}, and its exchangeability and CLT established in Theorems~\ref{thm:delta_identical} and \ref{thm:tau_clt}. Combination of the GDF and consistent high-dimensional clustering technique leads to our ClusMIP procedure (Algorithm~\ref{algo:clusmip}), which controls the FPR in both finite- and large-sample (Theorem~\ref{thm:clusmip_fpr}). Based on the simulations in Section~\ref{sec:simulation}, the ClusMIP procedure demonstrates a clear advantage over the existing  methods (DF(LASSO) and MIP) in terms of improved detection power, leading to considerably enhanced prediction power (lower MSE in prediction) and variable selection performance (higher empirical probability in selecting a submodel containing the true model) after removal of detected outliers. Specifically, the detection power of ClusMIP(SLASSO) is consistently performs best, varying between 0.90 and 1.00. In contrast, under-performance is seen for DF(LASSO) and MIP, where the detection power reaches levels between 0 and 0.20. Moreover, our simulation studies demonstrate that ClusMIP applied to all  model selectors identifies notably more outliers than existing methods, which impacts the selection of true predictors, further highlighting its advantage over the the exist methods. 

Applying the methods to the pain data, we find that the identified outliers from the ClusMIP procedure uniformly exhibit a tractable pattern corresponding to low pain ratings (Figure~\ref{fig:real_data_analysis_detection}(A)). This  substantiates the discrepancy between low and intense pain processing conditions. In contrast, those identified by DF(LASSO) show no recognizable pattern. The reduced model, obtained by removing observations detected by ClusMIP(MCP), is scientifically interpretable, containing unique features (i.e., the VAN network) that regulates the neuronal responses from the remaining high pain scores (Figure~\ref{fig:real_brain_plot}). 
In addition, the impact of excluding detected outliers on variable selection and coefficient estimation is clearly shown for all model selectors and detection procedures. Specifically, notable variation in the average model size and normed coef{}ficient estimates (at least $30\%$ change) are seen (Figure~\ref{fig:real_plot_aoas_revision_effects_detection_clusmip}(A) to (C), Figure~\ref{fig:real_plot_aoas_revision_effects_detection_dflasso}(A) and (B)). Remarkably, eliminating the outliers detected by ClusMIP(MCP) leads to the highest prediction power attained among all competing methods (Figure~\ref{fig:real_plot_aoas_revision_effects_detection_clusmip}(D) and (E)), with corresponding MSE  $26.62\%$ lower than those obtained using DF(LASSO) (Figure~\ref{fig:real_plot_aoas_revision_effects_detection_dflasso}(C) and (D)). Since MCP selects sparsest submodel with plausible scientific interpretation and highest prediction performance after removal of outliers identified by ClusMIP(MCP), it is recommended among all the competing proposals for this application.

\bibliography{References}
	
\end{document}